\documentclass{article}
\usepackage[margin=2.54cm]{geometry}
\usepackage{amsmath}
\usepackage{amsfonts}
\usepackage{booktabs}
\usepackage[utf8]{inputenc}
\usepackage{svg}
\usepackage{hyperref}
\usepackage{xcolor}
\usepackage{float}
\usepackage{subfigure}

\usepackage{graphicx}
\newcommand{\comment}[1]{}

\title{Containment strategies after the first wave of COVID-19\\
using mobility data}
\author{Martijn G\"osgens$^1$ \and Teun Hendriks$^1$ \and Marko Boon \and Stijn Keuning \and Wim Steenbakkers \and Hans Heesterbeek$^*$ \and Remco van der Hofstad$^*$ \and Nelly Litvak$^*$}
\date{\today}

\usepackage[backend=bibtex]{biblatex}
\begin{document}
\maketitle

\begin{abstract}
In their response to the COVID-19  outbreak, governments face the dilemma to balance public health and economy. Mobility plays a central role in this dilemma because the movement of people enables both economic activity and virus spread. We use mobility data in the form of counts of travellers between regions, to extend the often-used SEIR models to include mobility between regions.

We quantify the trade-off between mobility and infection spread in terms of a single parameter, to be chosen by policy makers, and propose strategies for restricting mobility so that the restrictions are minimal while the infection spread is effectively limited. We consider restrictions where the country is divided into regions, and study scenarios where mobility is allowed within these regions, and disallowed between them. We propose heuristic methods to approximate optimal choices for these regions.

We evaluate the obtained restrictions based on our trade-off. The results show that our methods are especially effective when the infections are highly concentrated, e.g., around a few municipalities, as resulting from superspreading events that play an important role in the spread of COVID-19.
We demonstrate our method in the example of the Netherlands. The results apply more broadly when mobility data is available.
\end{abstract}

\section{Introduction}

The pandemic of COVID-19, caused by the coronavirus SARS-CoV-2, has, by mid-October 2020, infected more than 40 million people in over 200 countries  and led to more than a million deaths. It is unlikely that the spread can be fully controlled in the near future and without the deployment of effective vaccines. Strategies are aimed at curbing exponential growth in case numbers and hospitalisations, predominantly to keep national health systems from becoming overburdened and to reduce infection pressure for people with a high risk of severe outcomes. Such strategies are limited to personal protection and hygiene, social distancing measures, reducing contacts and mixing/mobility. 
Although the virus is present globally, all countries implement their own strategies and sets of measures. 

At any given moment in the outbreak, there is a mix of countries and regions where the virus is temporarily under control, countries where the epidemic is decreasing and countries where the epidemic is increasing. 
After an initial peak in cases, countries remain at risk for second and subsequent peaks, even when no cases are reported in the country for long periods of time. As in principle everybody is susceptible to some degree, not reaching herd immunity after the initial wave of infection leaves a large susceptible population that can sustain subsequent outbreaks \cite{anderson2020will}. These new outbreaks can be triggered by infected individuals entering the country from outside, as a result of increased global mobility. Nationally, sustained transmission at relatively low levels can lead to new large (exponentially growing) outbreaks after the initial peak because control measures are relaxed or behaviour changes with respect to (social, temporal and spatial) mixing and personal protection/social distancing. Mixing increases the number of new contact opportunities that an infected individual has in the population and reduced effectiveness of personal protection and social distancing increases the probability per contact of transmission. Combined, these effects can lead to more transmission. Increased mixing not only reflects larger groups of individuals, but also reflects contacts with individuals from a larger geographic range, allowing infected individuals to have contacts with people from regions where infection pressure may hitherto be (very) low, causing clusters of cases in new areas. 

Mobility between areas plays a potentially important role in increasing transmission, but measures aimed at restricting mobility also have a potentially large controlling influence. Where, in the initial wave of infection, countries to a large extent imposed national mobility restrictions, the containment strategies for preventing subsequent waves of infection can perhaps be achieved by more regional or local mobility restrictions. This has the advantage of reducing the social and economic burden on society, but also has the risk that the restrictions may not be sufficiently effective and need to be scaled-up after all to a national level at some later point in time. It is, however, unclear how one could gauge the effectiveness of regional restrictions based on realistic mobility patterns specific to the country, balancing trade-offs between mobility and transmission. It is also unclear how large a `region' should be for effective containment and how different choices for recognizable regions (for example, administrative regions such as provinces, large cities, or postal code regions). 

In this paper, we provide a framework to evaluate the effectiveness of regional strategies aimed at restricting mobility, allowing for a range of choices of how regions are characterised, using the Netherlands as a case study. This is essential to be able to determine the scale at which interventions can be effectively imposed or lifted and addresses one of a range of key modelling questions for COVID-19 and future pandemic outbreaks \cite{thompson2020key}. We base the framework on actual mobility patterns in the Netherlands. We distinguish between extreme situations where infection is distributed evenly between areas and situations where infection is highly concentrated in a restricted area, for example as a result of a superspreading event. We show that regions defined on the basis of mobility patterns provide better strategies than regions based on administrative characteristics, and that focusing on administrative regions therefore leads to sub-optimal strategies. We also quantify and explore the non-linear relation between mobility and outbreak size for a range of choices of trade-off between mobility and transmission.

\section{Methodology}
In this section, we describe the overview of our approach. Given a certain set of regions, we envision a situation where mobility is allowed within the region, but mobility between the regions is not allowed. The main aim of this paper is to devise regions that allow for as much mobility as possible, yet restrict infections as much as possible. For this, we need to strike a careful balance between mobility and infections, which we formalize in terms of a trade-off parameter that policymakers need to impose. 

This section is organised as follows. We start in Section \ref{sec-strategies} by specifying the kind of mobility strategies that we consider in this work. Next, in Section \ref{sec:objective}, we introduce a way to quantify the performance of such strategies, by formulating the trade-off between mobility and infection spread as an explicit optimization problem over the various choices of regions described in Section \ref{sec-strategies}.
This trade-off is described in terms of the number of infections, for which we rely on an SEIR model that we introduce in Section \ref{sec-SEIR}, and mobility. In our SEIR model, the infections are described in terms of compartments that correspond to particular regions, and the infection is spread between regions by mobility between them. Our SEIR model takes such mobility into account, and relies on mobility information originating from telecommunication data. 

The optimization problem that formalizes our trade-off between mobility and infection containment is inspired by community-detection algorithms, and is complex to solve explicitly. In Section \ref{sec:opt}, we describe how to rigorously and heuristically analyze such problems. 
In particular, we provide heuristics that generate strategies with high performance. We close this section by describing how the various divisions in regions can be evaluated in Section \ref{sec:evaluation}.



\subsection{Strategies for mobility restrictions}
\label{sec-strategies}
We consider mobility restriction strategies of the following kind: given a division of the country into regions, we consider the scenario where movement is allowed {\em within} these regions, and disallowed {\em between} these regions.
We represent the country by a set of {\em atomic areas} $\mathcal{A}$. These areas are considered the smallest possible geographic units between which it is feasible to enforce mobility restrictions.
Then, a region is represented by a subset $D\subseteq\mathcal{A}$ between which mobility is allowed. Finally, a division is represented by a partition $\mathcal{D}=\{D_1,\dots,D_{|\mathcal{D}|}\}$. In our use case, we consider Dutch municipalities to be atomic areas.


Many administrative divisions that might serve as examples for regions $\mathcal{D}$ already exist. For example, we could use the divisions of the Netherlands and its municipalities into its 12 provinces, or 25 so-called security regions. An advantage of using such divisions is that they are already known so that it may be easier to communicate, and thus enforce, mobility restrictions based on them to the broad public. However, a disadvantage is that these divisions have been historically determined by decisions of governance, and thus their borders do not necessarily effectively reflect the actual movement of people throughout the country. As a result, mobility restrictions based on them may not be the most effective.

An illustrative example is the province of Flevoland (equal to the security region Flevoland). Almere, the most populous city of Flevoland, lies close to Amsterdam, where many of Almere’s citizens work. Our mobility data shows that more than 90\% of the mobility leaving Almere also leaves the province of Flevoland, as can be seen in Figure \ref{fig:mobility}. Therefore, choosing the division into provinces or security regions would disproportionately affect the people of Almere in terms of mobility. In Section~\ref{sec:opt}, we provide a method to obtain divisions based on the mobility data, by applying community detection methods. In Figure \ref{fig:mobility},  we see that divisions obtained by this method do consistently place Almere in the same region as Amsterdam.

Another disadvantage of using existing administrative divisions is that these are, by their very definition, inflexible and can hence not be tailored to the specific epidemiological situation. In Section~\ref{sec:opt}, we provide a method to obtain divisions that do take epidemiological information into account.

\subsection{Objective}
\label{sec:objective}
On the one hand, freedom of movement has both economic and intrinsic value. On the other hand, it also facilitates the spread of the disease. In essence, the problem for control is to find a trade-off between mobility and  infection containment, given the epidemiological characteristics and normal mobility patterns. In this section, we provide a way to {\em formalize} this trade-off to allow its characterisation.

\paragraph{Trade-off parameter.}
To formalize the trade-off between public health and mobility, we introduce a trade-off parameter $\gamma$. This parameter can be interpreted as the number of movements between areas that we are willing to restrict in order to prevent the occurrence of a {\em single} further infection. A higher value for $\gamma$ thus favors more severe restrictions. The choice for this trade-off parameter reflects societal values and is hence a political choice that should be made by politicians or policy makers. Therefore, we refrain from giving advice about a specific suitable value, but instead provide a method that advises a strategy given a choice for $\gamma$.

\paragraph{Time horizon.}
Suppose that a certain set of mobility restrictions is in place. We consider some time horizon $H$ representing the number of days that the restriction will be in place, and count the number of infections and movements before this horizon. This time horizon should thus not be too long, as it coincides with the duration that restrictions are in place and sometimes one needs to quickly respond to changes in the infection spread. Due to the delay to go from a contact moment to an infection, the time horizon should also not be too short, as otherwise the effect of the imposed restrictions cannot reasonably be observed. In this work, we use a time horizon of $30$ days as an example.

At the end of this time horizon, a new division into regions may be chosen based on the status of the epidemic at that moment in time, thus allowing for a dynamical update of the strategy of mobility restrictions.

\paragraph{Objective function.} The above considerations lead to the objective function
\begin{equation}\label{eq:tradeoff}
Q_{\gamma,H}\left(\mathcal{D}\right)=\mathcal{M}\left(\mathcal{D};H\right)-\gamma\cdot G\left(\mathcal{D};H\right),
\end{equation}
where $\mathcal{M}\left(\mathcal{D};H\right)$ and  $G\left(\mathcal{D};H\right)$ represent the number of movements and infections, respectively, that occur before the time horizon $H$, given a division $\mathcal{D}$. Note that in this formulation the current status of the epidemic, i.e., the number of infectious and susceptible people at the start of the period $H$, is included because the value of $G\left(\mathcal{D};H\right)$ depends on  that initial status. 
The goal of this study is to provide a methodology that, given $\gamma$ and $H$, finds a division $\mathcal{D}$ with high $Q_{\gamma,H}\left(\mathcal{D}\right)$.  

\paragraph{Estimating mobility.}
The company Mezuro has a platform that produces mobility patterns based on telecom data. It provides information about the average number of people that move between the different municipalities in a given time period, in our case study this is the period from March 01 2019 up to and including March 14 2019.  Given a division, the mobility between two municipalities in the same region is estimated by the average mobility observed over these two weeks. The mobility between two municipalities in different regions is assumed to be zero, corresponding to full compliance to the restrictions. By using a daily average, we lose the difference between weekdays and weekends. Also, the period on which the average in our mobility data is based is obviously before the COVID-19 outbreak. We interpret this data as being the mobility benchmark: this is what people {\em would} travel if COVID-19 were not present. Thus, any mobility that is less arises through the governmental and societal measures that are put upon our society.

\paragraph{Estimating infections.}
Infection dynamics are described by a modified version of a commonly used compartmental SEIR model \cite{Diekman2013Tools}, where people can be either susceptible (S), exposed (E), infectious (I) or removed (R). Exposed people are infected but become infectious after a latent period. Removed people are recovered from the disease, assuming live-long immunity, or passed away caused by the disease. Differential equations govern the rates of infection spread between and within the different compartments. The dynamics are stochastic and discrete. In our model, we capture the spatial component of infection spread, population distribution, and connectivity between different regions. This model is explained in the next section.

\subsection{Region-based epidemiological model}
\label{sec-SEIR}
We propose a modified SEIR model that incorporates mobility, and we explain next.

\begin{figure}[!ht]
    \centering
    \includegraphics[width=0.6\textwidth]{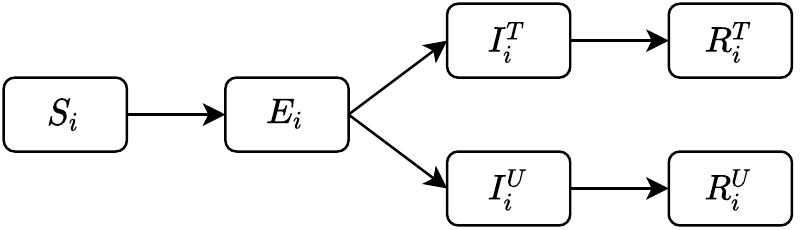}
    \caption{The SEI$_2$R$_2$ compartmental model within municipality $i$. $U,T$ stand for `Untested' and `Tested'. The rate of new infections $S_i \rightarrow E_i$ consists of a local part within municipality $i$ and a non-local part which depends on infectious people from other municipalities via mobility to and from these municipalities.}
    \label{fig:modeldiagram}
\end{figure}

Firstly, to take into account the spatial component and population distribution, we divide the total population $N$ into $n$ smaller groups, designating spatial distinct areas with population size $N_i$, $i\in [1,2,\ ...,n]=\mathcal{A}$, such that $N=\sum_{i=1}^n N_i$. Such an approach has been commonly used in compartmental models \cite{tizzoni2014use,sattenspiel1995structured}. We call the groups {\em atomic areas}. Using municipalities in the Netherlands would set $n=380$ (based on 2018). To potentially model control and behavioural effects on mobility better for SARS-CoV-2, we distinguish between tested and untested individuals in infectious and removed compartments, indicated by the superscripts $T$ and $U$, respectively. Fig. \ref{fig:modeldiagram} shows the different compartments. Because there are two infectious and two removed compartments, we abbreviate it to SEI$_2$R$_2$. Others use a similar approach to modeling SARS-CoV-2 \cite{Li2020, liu2020model,teslya2020impact}.


To simulate connectivity between different atomic areas, we introduce the mobility parameter $M_{ij}(t)$, which is interpreted as the number of individuals from atomic area $j$ who visit atomic area $i$ at time $t$. If areas are taken as municipalities, then we can use Mezuro mobility data as a proxy for $M_{ij}$. People are assumed to always return home at the end of every day, i.e. visits to other locations are assumed to be brief.

In the SEIR model, an important parameter is the transmission rate $\beta$. We assume the number of contacts per person per unit time to be independent of the population size, which is called standard incidence. In this case, $\beta$ is the product of the contact rate $c$ (contacts per person per unit time) and the transmission probability of the virus $\varepsilon$. Our model aims to capture the reduction in contacts a person has when their mobility is restricted. A lower contact rate translates into a lower transmission rate of the virus and ultimately. 

To achieve this in our mobility setting, we split $c$ in local contacts within one's atomic area $j$, and non-local contacts resulting from travel to any other another atomic area $i$, as described by the $M_{ij}$. We assume local contacts to be a fraction $p$ of the overall average contact rate; $c_{\rm loc}=pc$. In the population of size $N$ the total number of meetings is $cN/2$, assuming a contact is only between two people, Of which a fraction $p$ is now account for locally. The remainder of contacts are made through travelling people who visit other areas and mix with the individuals present there. There are $\sum_{i,j\in\mathcal{A}}M_{ij}$ travelling people. We calculate the contacts per travelling person per unit time using:
\[c_{\rm mob}=(1-p)c\frac{N}{2\sum_{i,j\in\mathcal{A}}M_{ij}},   
\]
 In a system where we impose no restriction on mobility, we have an overall average contact rate equal to $c$, but when mobility between regions is restricted, the overall average contact rate decreases. 

We define $\beta_{\rm loc}=\varepsilon c_{\rm loc}$ and $\beta_{\rm mob}=\varepsilon c_{\rm mob}$. We next explain how $\beta_{\rm loc}$ and $\beta_{\rm mob}$ can be incorporated in the SEI$_2$R$_2$ model. New infections arise by three different mechanisms. (1) locally, infectious people infect susceptible within their atomic area $i$, (2) susceptibles from area $i$ visit area $j$ and get infected, or (3) infectious people from area $j$ visit area $i$ and infect susceptible inhabitants of region $i$. We assume that tested infectious people minimize their contacts to only local contacts, so that they do not play a role in spreading the virus to other areas.

This results in the following mean-field differential equations for the dynamics of the model:

\begin{align*}
\allowdisplaybreaks
\frac{dS_i\left(t\right)}{dt}&=-\beta_{\rm loc}\frac{S_i\left(t\right)}{N_i}\left(I_i^\text{T}\left(t\right)+I_i^\text{U}\left(t\right)\right)-\beta_{\rm mob}\sum_{j\in D} \left( S_i(t)\frac{M_{ji}}{N_i}\frac{I^\text{U}_j(t)}{N_j} +I^\text{U}_j(t)\frac{M_{ij}}{N_j}\frac{S_i(t)}{N_i}  \right), \\
\frac{dE_i\left(t\right)}{dt}&=\beta_{\rm loc}\frac{S_i\left(t\right)}{N_i}\left(I_i^\text{T}\left(t\right)+I_i^\text{U}\left(t\right)\right)+\beta_{\rm mob}\frac{S_i(t)}{N_i}\sum_{j\in D}  \frac{I^\text{U}_j(t)}{N_j}\left (M_{ji} +M_{ij}\right)-\frac{E_i\left(t\right)}{\nu},\\
\frac{dI_i^\text{T}\left(t\right)}{dt}&=a\frac{E_i\left(t\right)}{\nu}-\frac{I_i^\text{T}\left(t\right)}{\omega},\\
\frac{dI_i^\text{U}\left(t\right)}{dt}&=\left(1-a\right)\frac{E_i\left(t\right)}{\nu}-\frac{I_i^\text{U}\left(t\right)}{\omega},\\
\frac{dR_i^\text{T}\left(t\right)}{dt}&=\frac{I_i^\text{T}\left(t\right)}{\omega},\\
\frac{dR_i^\text{U}\left(t\right)}{dt}&=\frac{I_i^\text{U}\left(t\right)}{\omega}.
\end{align*}

Here atomic area $i$ is by convention part of region $D$. We have the following parameters: $a$ is the fraction of people which gets tested,  $\nu$ is the latent period, and $\omega$ is the duration of the infectious period. Estimates of all parameters can be found in Table~\ref{table-1}. Summation over $j\in D$ means we only allow travel from atomic area $i$ to and from atomic area $j$ if it is within the same partition $D$ as atomic area $i$, where $D\in \mathcal{D}$.

\begin{table}[]
\begin{tabular}{@{}llll@{}}
\toprule
Name                                  &                 & Value  & Source \\ \midrule
Fraction tested                       & $a$               & 1/15   & estimated        \\
Fraction local contacts               & $p$               & 1/2    &   estimated     \\
Infectious period                     & $\omega$        & 5 days      & Deng et al \cite{deng2020estimation}       \\
Latent period                         & $\nu$           & 4 days     &  Shorter than incubation\\ &&&period \cite{world2020transmission,linton2020incubation}      \\
Basic reproduction number             & $\mathcal{R}_0$ & 2.5    &  Li et al \cite{li2020early}      \\
Prevention measure efficacy          &                 & $50\%$ &  estimated      \\
Effective reproduction number (with prevention measures) &   $\mathcal{R}_{\rm eff}$         & 1.25   & 50\% of $\mathcal{R}_0$  \\ 
Transmission probability per contact with infectious person     &  $\varepsilon$               &  0.0238      &  NGM and $\mathcal{R}_{\rm eff}=1.25$       \\
Average contact rate (unique persons) & $c$               & 13.85  &  Mossung et al \cite{mossong2008social}      \\
Transmission rate via local contacts  & $\beta_{\rm loc}$  & $0.165$   & $\beta_{\rm loc}=c_{\rm loc}\varepsilon$               \\
Transmission rate via mobility related contacts  & $\beta_{\rm mob}$   &$0.141$  & $\beta_{\rm mob}=c_{\rm mob}\varepsilon$             \\
\bottomrule
\end{tabular}
\caption{Values of the parameters in our epidemic model.}
\label{table-1}
\end{table}

It is estimated that the basic reproduction number $\mathcal{R}_0$ of COVID-19 is in the range $[2, 3]$ in a population without additional measures such as social distancing, sneezing in your elbow and regular hand-washing \cite{li2020early}. In our model, these measures are not modelled explicitly, but they are accounted for by using an effective reproduction number $R_{\rm eff}$ without mobility restrictions smaller than $2$. This corresponds to a situation where all mobility is allowed combined with preventive measures being in place, resulting in a lower $\mathcal{R}_{\rm eff}$ and hence a lower $\varepsilon$. For exposition, use a value $\mathcal{R}_{\rm eff}=1.25$. For such a value, the effective $\mathcal{R}_{\rm eff}(\mathcal{D})$ including the mobility measures will be smaller than 1 when no mobility is allowed, so that an effective choice of mobility restrictions can in fact significantly reduce and contain the infection spread.

The choice of $\varepsilon$ is based on this effective reproduction number $\mathcal{R}_{\rm eff}$ in the situation where there are no restrictions, so that all mobility is allowed. $\mathcal{R}_{\rm eff}$ is calculated as the dominant eigenvalue of the Next Generation Matrix (NGM) \cite{diekmann2010construction}. This calculation can be found in Section \ref{sec-sensitivity} in the supplementary material. For $\mathcal{R}_{\rm eff}=1.25$, we have $\varepsilon = 0.0238$.

\paragraph{Initialization of the model.}
To run our model, we further need to decide where the infections are located at the start of the time horizon. We call this the {\em model initialization}. We initialize the model by choosing the number of people that reside in each compartment for each municipality. We distinguish two kinds of initializations: \textit{synthetic initializations} and \textit{historical initializations}, the latter based on data from RIVM\footnote{The RIVM is the Dutch institute for public health. The data is available at \url{https://data.rivm.nl/covid-19/}.}.
The synthetic initializations provide insight in the spread of the infection based on our SEI$_2$R$_2$ model, and the performance of the various regional sub-divisions, while the historical initializations show to what extent these insights generalize to practical settings.

For the synthetic initializations, we start by choosing the number of exposed inhabitants for each municipality and set the remainder of the populations to susceptible.
In a deterministic approach, these numbers do not need to be integer. In a stochastic setting, non-integer values are rounded (see Section \ref{sec-comp-deter-stoch} in the supplementary material). For the spread of the initial exposures, we consider two extreme cases. In the first case, a {\em fixed fraction} of each municipality's population is set to exposed. In the second case, we place {\it all} exposed individuals in one municipality. After the exposures are set,  we simulate the model for 10 days so that sufficiently many exposures have led to infections and new exposures, because in practice this is the point when the outbreak is detected and the measures are introduced. 
This leads to two synthetic initializations that we will refer to as \textit{evenly distributed} and \textit{concentrated}, respectively.
The evenly distributed initialization mimics a widespread outbreak while the concentrated initialization represents a superspreader event. See Section~\ref{sec-init-details} in the supplementary material for details. 

\begin{figure}[!ht]
  \centering
  \includegraphics[width =\textwidth]{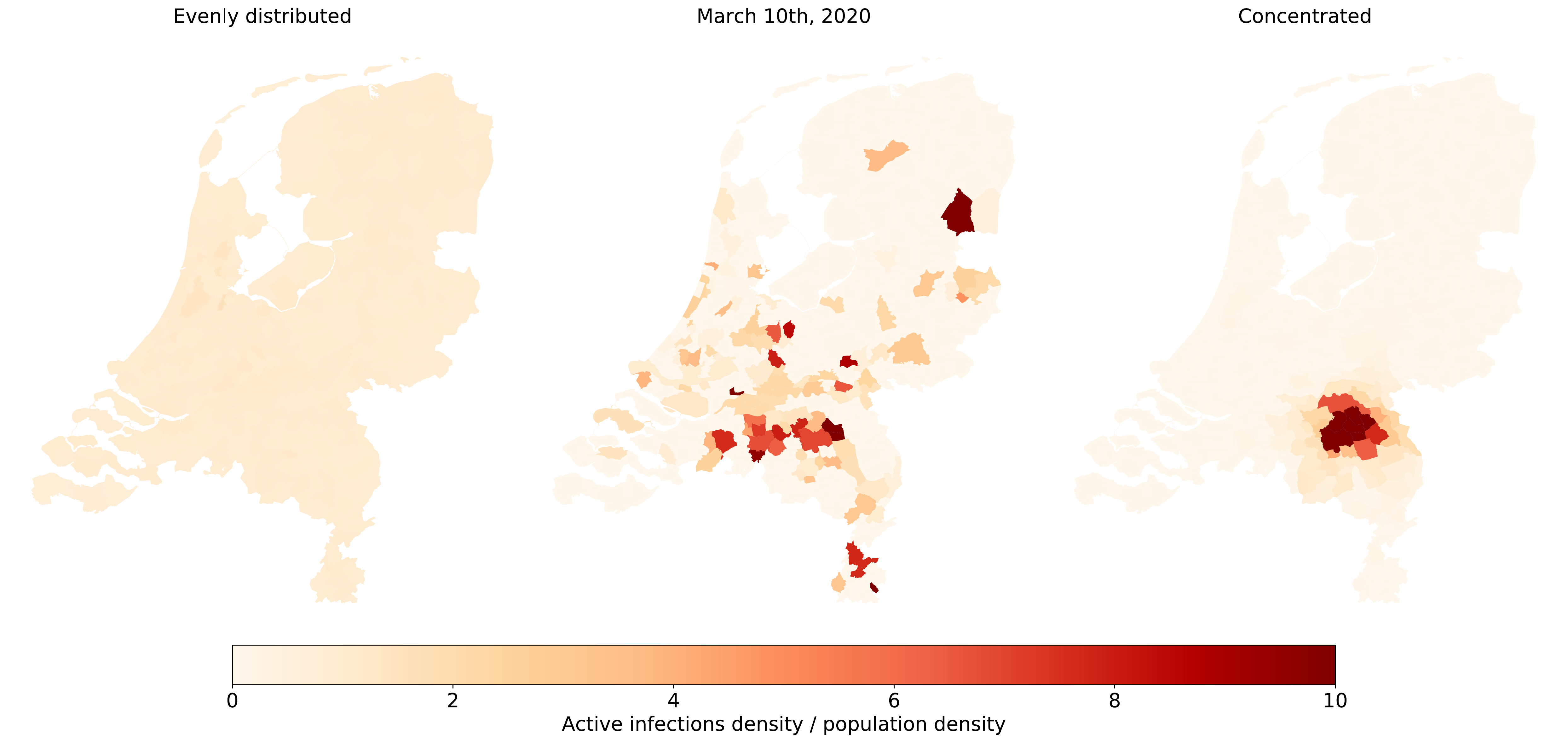}
  \caption{Difference in concentration of infectious people after initialization. The model is initialized with either evenly distributed, historical, or  concentrated distributions of active infections. For the synthetic initializations, 1000 people are exposed and 10 days are simulated. The active infection density equals the number of infections within the municipality divided by the national total number of infections. The population density equals the municipality population divided by the total population.}
  \label{fig:initializations}
\end{figure}

For the historical initialization, we assume that daily updates are available on the cumulative number of confirmed infections in each municipality.
Based on these daily cumulative numbers, the number of active tested infections can be estimated by the difference between the cumulative numbers at time $t$ and one infectious period before that. The number of untested infections can be found using the assumption that a fraction $a$ of the population is tested. The number of exposed individuals can be found by looking ahead one latent period. A more detailed description of these initializations can be found in the supplementary material (see Section~\ref{sec-init-details}. Figure~\ref{fig:initializations} shows how the infections are distributed throughout the country in each of these initializations. 

\begin{figure}[!ht]
 \centering
 \includegraphics[width=0.7\textwidth]{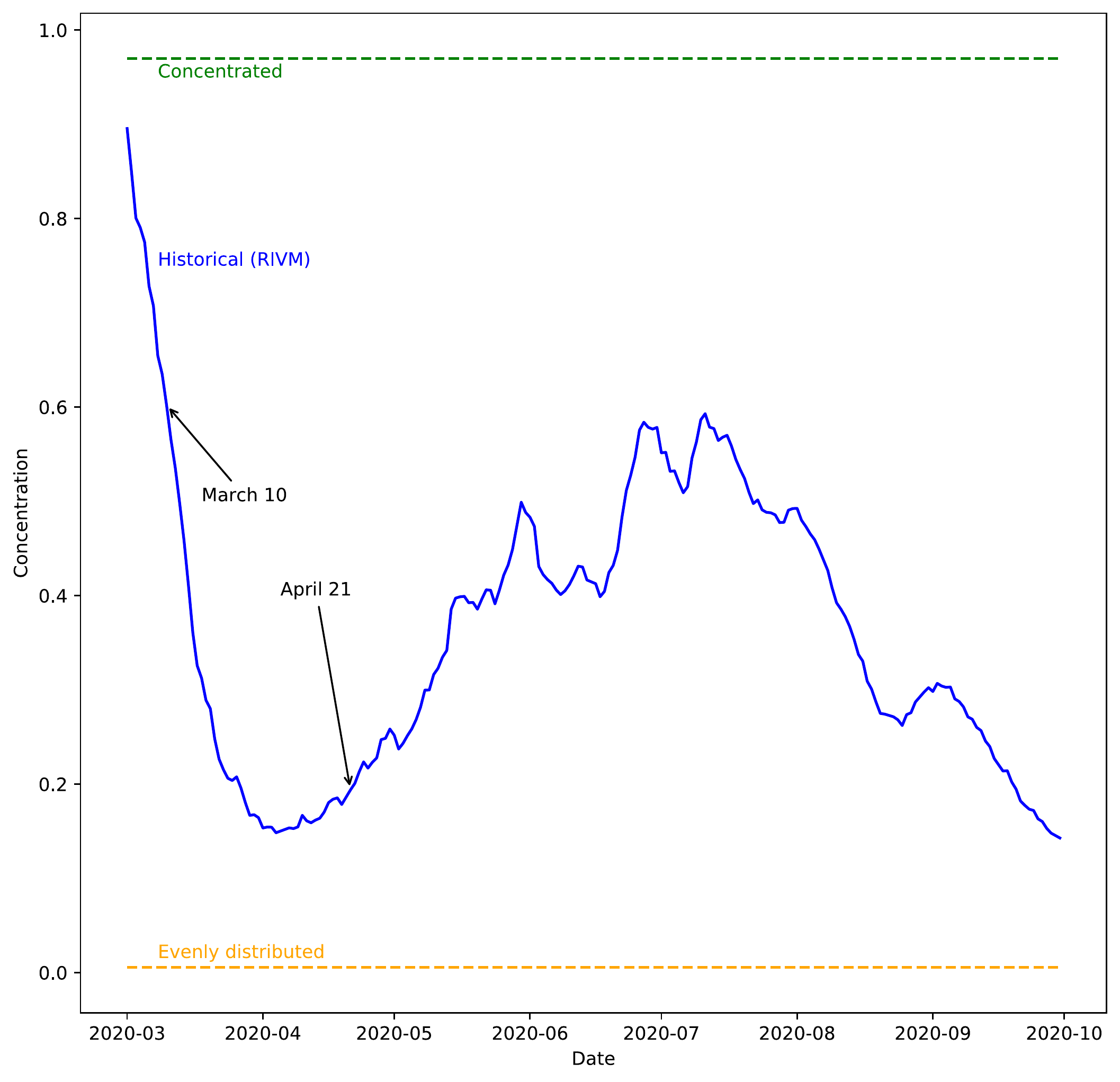}
 \caption{The concentration values corresponding to distributions of infections according to the RIVM data over time. The dotted lines show the concentration levels of the synthetic initializations. Data obtained from \url{https://data.rivm.nl/covid-19/}.}
 \label{fig:concentration}
\end{figure}

To quantify the concentration of the infections, we compare the distribution of the infections over the municipalities to the distribution of the population over the municipalities. For this, we introduce an entropy-type measure. Let $p_i^\text{inf}$ denote the fraction of infectious individuals that live in municipality $i$ and let $p_i^\text{pop}$ denote the fraction of the population that lives in municipality $i$. If the infections would be distributed evenly, then these two distributions would be equal. We measure the concentration of the infections by $1-\mathrm{e}^{-D_{\rm KL}}$, where $D_{\rm KL}$ denotes the Kullback-Leibler divergence from the infection distribution $\left(p_i^\text{inf}\right)_{i\in\mathcal{A}}$ to the population distribution $\left(p_i^{\mathrm{pop}}\right)_{i\in\mathcal{A}}$. 
Note that this measure does not depend on the total number of infections. When each municipality has the same percentage of infected individuals, the concentration will equal $0$. On the other hand, when all infections are located in the same municipality, then it will be close to $1$.
This allows us to quantify the concentration of the infections for our initializations on a scale from $0$ to $1$. Synthetic initializations are close to $0$ and $1$, respectively, while historical initializations have intermediate values.
The concentration values of initializations based on the historical data obtained from the the Dutch National Institute for Public Health and the Environment (RIVM) are shown in Figure~\ref{fig:concentration}.
In the beginning of the epidemic, there were a few local outbreaks, resulting in a high concentration. As the outbreak became more widespread, the concentration went down, reaching a minimum at the beginning of April. The period after April is characterized by small local outbreaks, leading again to higher concentration values.

\subsection{Optimization}
\label{sec:opt}
The quantification of $Q_{\gamma,H}\left(\mathcal{D}\right)$ in Equation \eqref{eq:tradeoff} is computationally intensive because it involves computation of the number of infections at the end of the time horizon, and the number of possible divisions $\mathcal{D}$ is enormous, it is not feasible to find the global optimum. Therefore, we will resort to heuristic optimization methods.
To do this, we will iterate the Louvain algorithm \cite{Blondel2008}, which can find partitions that are local optima with respect to the following manipulations \cite{Traag2019}: moving a single element from one set to the other and merging together two sets.
This algorithm is able to optimize a wide range of functions over partitions \cite{Prokhorenkova2019}.
However, each re-evaluation of the score involves running a simulation in our setting, this method is computationally expensive if the initial division of atomic regions $\mathcal{A}$ consists of many small elements. For example, at the finest sub-division possible in our setting, municipality-level, the optimization takes more than 24 hours, based on the 380 municipalities in the Netherlands. 
To improve on this, we define coarser sub-divisions as starting points for the optimization procedure. Instead of letting Louvain operate at the level of single municipalities, we will start from an initial division of the municipalities into sub-regions that will not be further divided by the algorithm. 
This way, Louvain is guaranteed to result in a division that cannot be improved upon by merging two regions or by moving one sub-region to another region \cite{Traag2019}.
There are multiple ways to choose such initial divisions. One can use existing administrative divisions such as the twelve provinces  of the Netherlands or its twenty five security regions. Such administrative divisions may already be known by the public, making it easier to enforce restrictions based on them. A disadvantage can be that such regions have not been defined with an outbreak of an infectious disease in mind and therefore do not pose natural boundaries to transmission. Neither have they been defined on the basis of economic activity. We can also use other criteria to find other initial divisions, for example based on behaviour of individuals that relates to transmission or to economic activity. Human mobility may be a good indicator for both of these. We give two criteria for obtaining such initial divisions.

\paragraph{Mobility regions.}
In network science, the objective of community detection is to partition the nodes of a network into groups that are more highly connected internally than externally \cite{Fortunato2010}.
Community detection has been applied to mobility networks \cite{Hossmann2011,Yildirimoglu2018}, resulting in divisions into regions that are coherent with respect to mobility.

Currently, the most popular community detection method is to optimize a quantity called {\em modularity}, which computes the weight inside the communities minus the expected weight for a random network with the same weights.
For mobility data of the kind that we rely on, the modularity of a division $\mathcal{D}=\{D_1,\dots,D_{|\mathcal{D}|}\}$ is given by
\[
\text{Modularity}_{\eta}\left(\mathcal{D}\right)=\frac{1}{M\left(\mathcal{A},\mathcal{A}\right)}\sum_{D\in\mathcal{D}}\left[ M\left(D,D\right)-\eta\frac{M\left(D,\mathcal{A}\right)M\left(\mathcal{A},D\right)}{M\left(\mathcal{A},\mathcal{A}\right)}\right],
\]
where $M\left(A,B\right)=\sum_{i\in A}\sum_{j\in B} M_{ij}$ is the mobility between the atomic areas in $A\subseteq\mathcal{A}$ and in $B\subseteq\mathcal{A}$, and $\eta$ is the resolution parameter, which controls the granularity of the found communities \cite{Fortunato2007}.
Larger values of $\eta$ result in divisions consisting of more regions.

\begin{figure}[!ht]
  \centering
  \includegraphics[width =\textwidth]{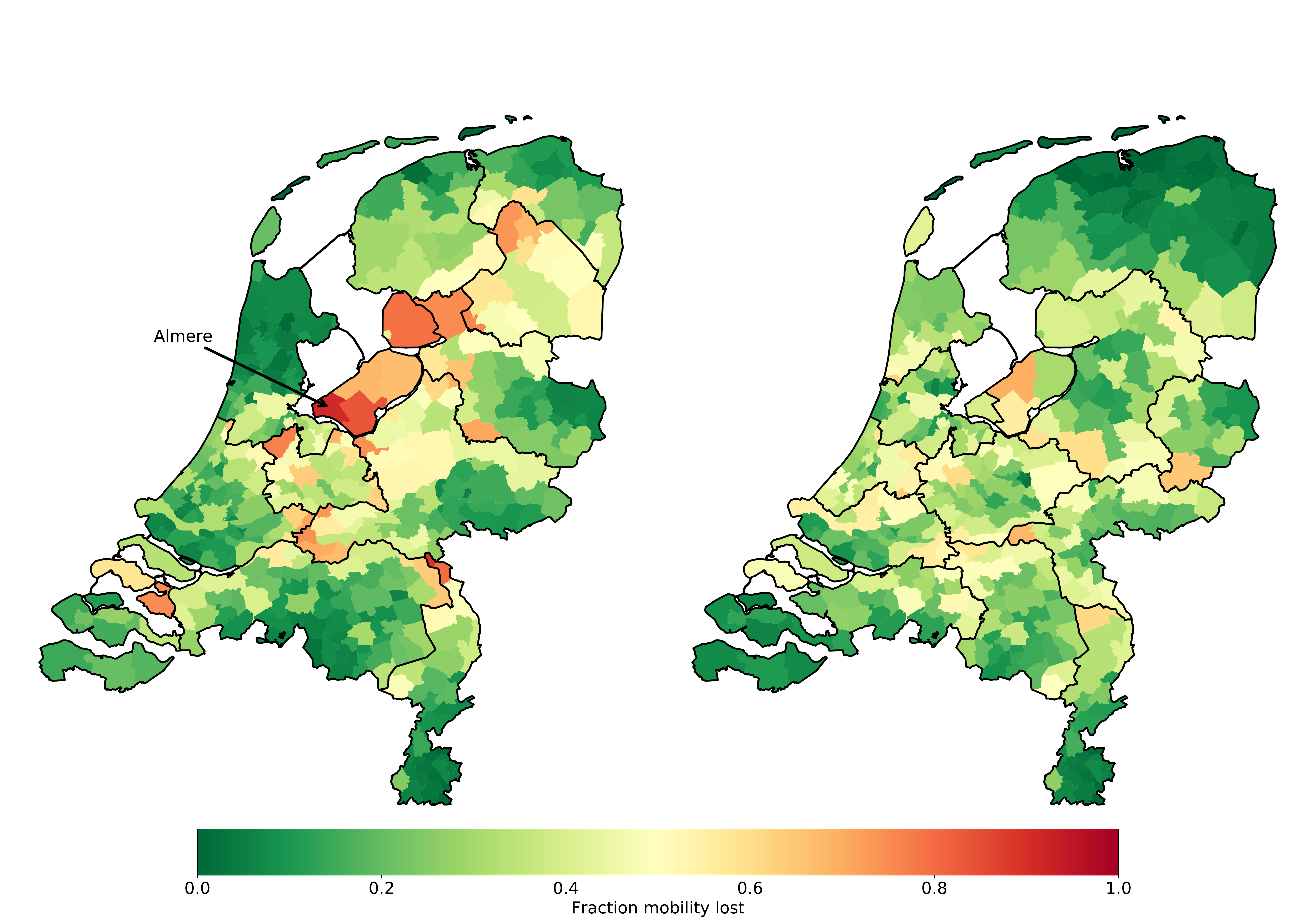}
  \caption{Each municipality is colored based on the percentage of mobility to destinations outside its region. Left: provinces. Right: mobility regions found by modularity optimization with $\eta=2$, chosen such that both divisions consist of 12 regions.}
  \label{fig:mobility}
\end{figure}

We iterate the Louvain algorithm  \cite{Blondel2008} to optimize modularity.
We will refer to regions obtained by this optimization as {\em mobility regions}.
We note that the number of mobility regions that result from applying this method for a given resolution parameter is in general not known beforehand.
Therefore, we apply this method for a variety of resolution parameter values to obtain divisions consisting of different numbers of mobility regions.
For example, by trial and error it was found that the choice $\eta=2$ resulted in a division into 12 mobility regions, so that this division has comparable granularity as the division into the 12 provinces.
A comparison of these two divisions based on mobility is shown in Figure~\ref{fig:mobility}.
We see that these mobility regions are indeed more coherent in terms of mobility.
In particular, Almere is in the same mobility region as Amsterdam, but in a different province. Mobility regions may reflect the economic and non-local transmission activity of the citizens in a better way. However, they cannot be tailored to the status of the epidemic.

\begin{figure}[!ht]
  \centering
  \includegraphics[width =\textwidth]{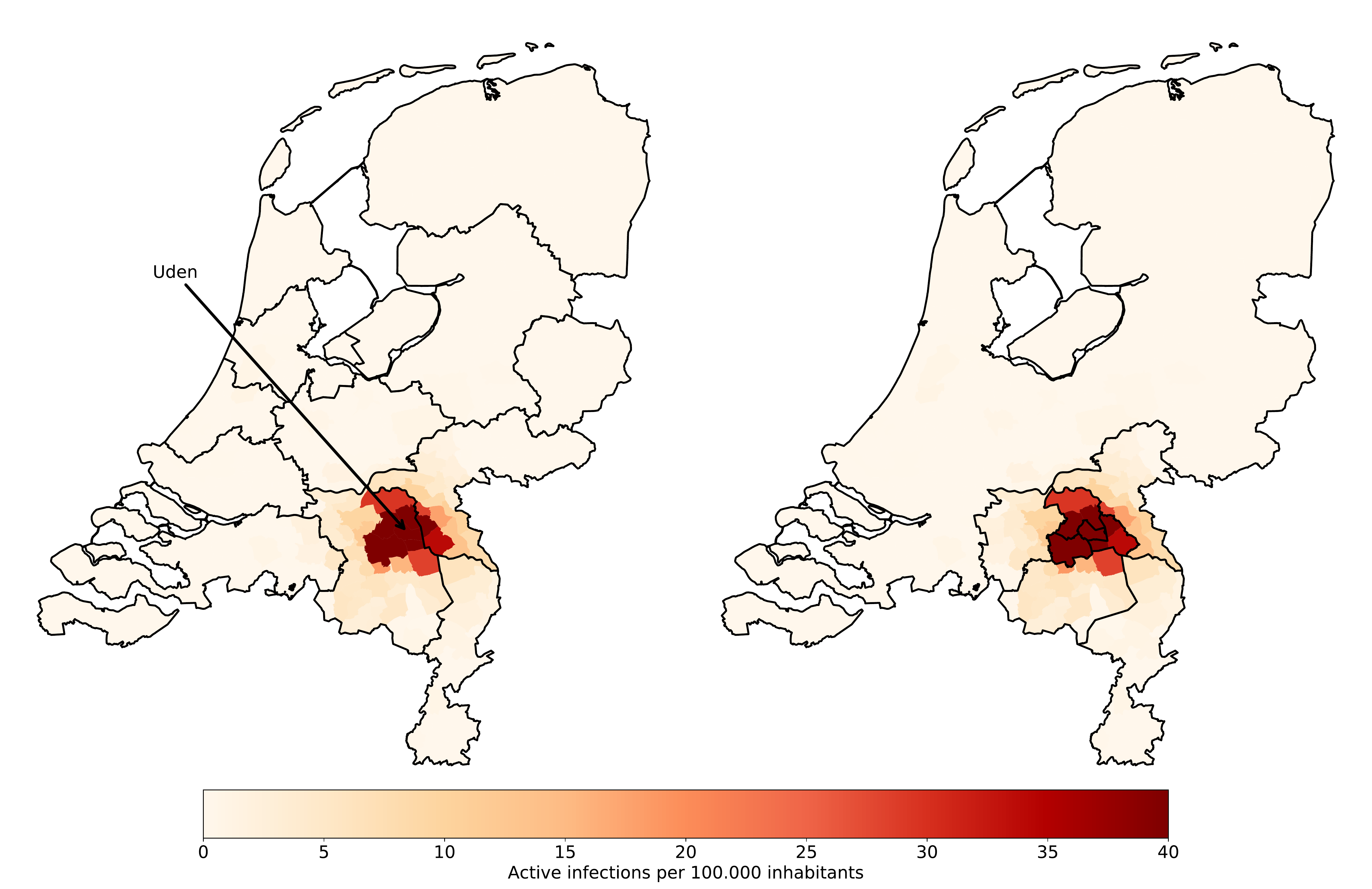}
  \caption{The mobility regions (for $\eta=2$) on the left versus the adaptive mobility regions (for $\zeta=75000$) on the right. The colors denote the number of active infections per capita for the concentrated initialization. For the colorized figure, we refer to the online version of this article.}
  \label{fig:mitigation_regions}
\end{figure}

\paragraph{Adaptive mobility regions.}
The running time of our optimization algorithm depends heavily on the number of sub-regions in our starting division.
Note that for the mobility regions, the resolution parameter controls the resolution of the division globally. However, high resolution is mostly needed around locations where a lot of infections occur.
We next introduce a modification to the modularity function of the previous section to obtain initial divisions that have a higher resolution near such critical locations.
Infections are due to meetings between infectious and susceptible individuals. Therefore, a good heuristic would aim for a division that separates infectious individuals from susceptible ones as much as possible. For a region $D\subset\mathcal{A}$, let $I\left(D\right),S\left(D\right)$ denote the number of infectious and susceptible individuals living in $D$, respectively. We maximize the function
\[
\text{AdaptiveModularity}_{\zeta}\left(\mathcal{D}\right)=\sum_{D\in\mathcal{D}}\left[ M\left(D,D\right)-\zeta\frac{I\left(D\right)S\left(D\right)}{N}\right],
\]
where $\zeta$ is a resolution parameter that, similarly to the $\eta$ parameter of modularity, determines the granularity of the obtained division. In particular, by varying $\zeta$, the method results in divisions with different numbers of regions. 
Despite the fact that $\zeta$ has a similar role to the $\eta$ of the mobility regions, it does have different dimension and differs in an order of magnitude.
Again, the Louvain algorithm is iterated for optimization.
We call the resulting regions {\em adaptive mobility regions} because the resolution is locally adapted to the state of the epidemic.
Figure~\ref{fig:mitigation_regions} compares a division into mobility regions to a division into adaptive mobility regions for the concentrated initialization. We see that the adaptive mobility regions indeed have a higher resolution around the critical area of the superspreader event in Uden.


\subsection{Evaluation of divisions}\label{sec:evaluation}
Given values for the trade-off parameter $\gamma$ and the time horizon $H$, we can compute the value $Q_{\gamma,H}\left(\mathcal{D}\right)$ to assess the quality of some newly obtained division $\mathcal{D}$. However, by itself this abstract value has no clear interpretation. We can obtain insight by comparing the quality of $\mathcal{D}$ to the quality of existing alternative divisions. These existing divisions are referred to as benchmark divisions. 
There exist administrative choices to divide the country into regions, such as provinces and security regions. We consider the following benchmark divisions for the situation of a widespread outbreak:
\begin{itemize}
    \item No restrictions (minimum restrictions);
	\item Disallow movement between provinces;
	\item Disallow movement between security regions;
	\item Disallow movement between municipalities (maximum restrictions).
\end{itemize}
Note that minimum restrictions maximize mobility and infections while these are both minimized by maximum restrictions.
For a superspreading initialization in a single municipality, we also consider the following alternative benchmark divisions:
\begin{itemize}
	\item Isolation of the municipality;
	\item Isolation of the security region of this municipality;
	\item Isolation of the province of this municipality.
\end{itemize}
Given a value for the trade-off parameter, we can assess whether the division that is obtained by optimizing the objective for this trade-off value indeed outperforms the benchmark divisions.

Finally, given a set of divisions $\mathcal{D}_1,\ldots,\mathcal{D}_k$ and a horizon $H$, we can plot $\mathcal{M}\left(\mathcal{D};H\right)$ and $G\left(\mathcal{D};H\right)$ for each division, resulting in plots such as in Figures~\ref{fig:evenlydistributed}, \ref{fig:concentrated}, \ref{fig:historical0310}, and~\ref{fig:historical0421}.
In these plots, when a division is plotted to the right of a benchmark division, it is more favorable in terms of mobility, while a lower vertical position indicates fewer infections. When a division is such that both are the case, we can say that a division \emph{dominates} the benchmark division: for any value of the trade-off parameter, it will be favored over the benchmark division.
Note that it is not possible to dominate the minimum and maximum restrictions benchmarks, since they achieve maximum mobility and minimum infections, respectively.

Given a set of divisions $\mathcal{D}_1,\ldots,\mathcal{D}_k$ and a choice for the trade-off parameter and the horizon, the division with the highest objective value can be chosen.
Obviously, the resulting division is not guaranteed to be the global optimizer of the objective as this would require comparing an enormous amount of divisions.
Therefore, the quality of such choice depends on the quality of the candidates  $\mathcal{D}_1,\ldots,\mathcal{D}_k$.
\section{Results}
For various initializations, we perform the  optimization described in Section~\ref{sec:opt} and simulate each of the strategies.
Then we evaluate the results based on the objective defined in Section~\ref{sec:objective}.
Throughout this section, we consider a time horizon of $H=30$ days.
For each initialization, we apply the optimization method to various initial divisions. The initial divisions will be the administrative divisions (provinces and security regions) and mobility-based initial divisions (mobility regions and adaptive mobility regions for various values of $\eta$ and $\zeta$).
Recall that we refrain from advising a specific value for the trade-off parameter since it is a political choice. However, in this section we will consider some values for $\gamma$, but these are intended only to demonstrate the methodology.
We will consider two values for $\gamma$: the value $\gamma^*$ for which maximum and minimum restrictions have equal objective value (i.e. $Q_{\gamma^*}(\mathcal{D}_{\min})=Q_{\gamma^*}(\mathcal{D}_{\max})$ where $\mathcal{D}_{\max}=\mathcal{D}_{\min}$ are the divisions corresponding to maximum, respectively, minimum restrictions) and twice this $\gamma^*$.
Note that this value $\gamma^*$ is uniquely defined since $Q_{\gamma,H}(\mathcal{D})$ is a linear function in $\gamma$.
We choose these values for the trade-off parameter rather than fixed constants because constant trade-off parameters may lead to minimum restrictions for one initialization and maximum restrictions for another. This choice of $\gamma^*$ is dependent on the initialization and ensures that there is a non-trivial division that outperforms both maximum and minimum restrictions. 
We start with the synthetic initializations and draw a few observations. Then, we see to what extent these observations generalize to the historical initializations.

\begin{figure}[!ht]
 \centering
 \subfigure[Evenly distributed]{
        \label{fig:evenlydistributed}
        \centering
        \includegraphics[width=0.45\textwidth]{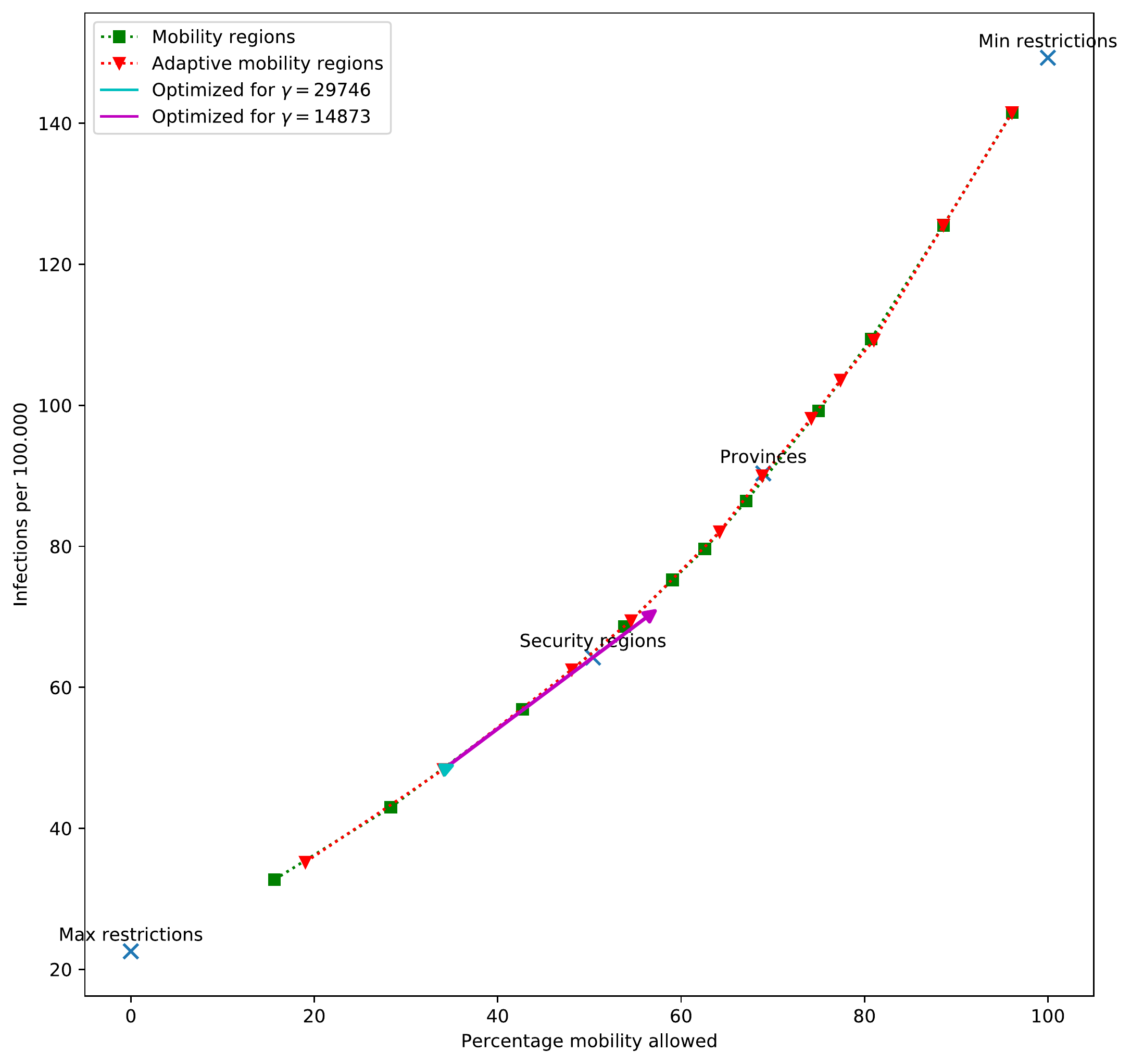}
    }%
    \qquad
    \subfigure[Concentrated]{
        \label{fig:concentrated}
        \centering
        \includegraphics[width=0.45\textwidth]{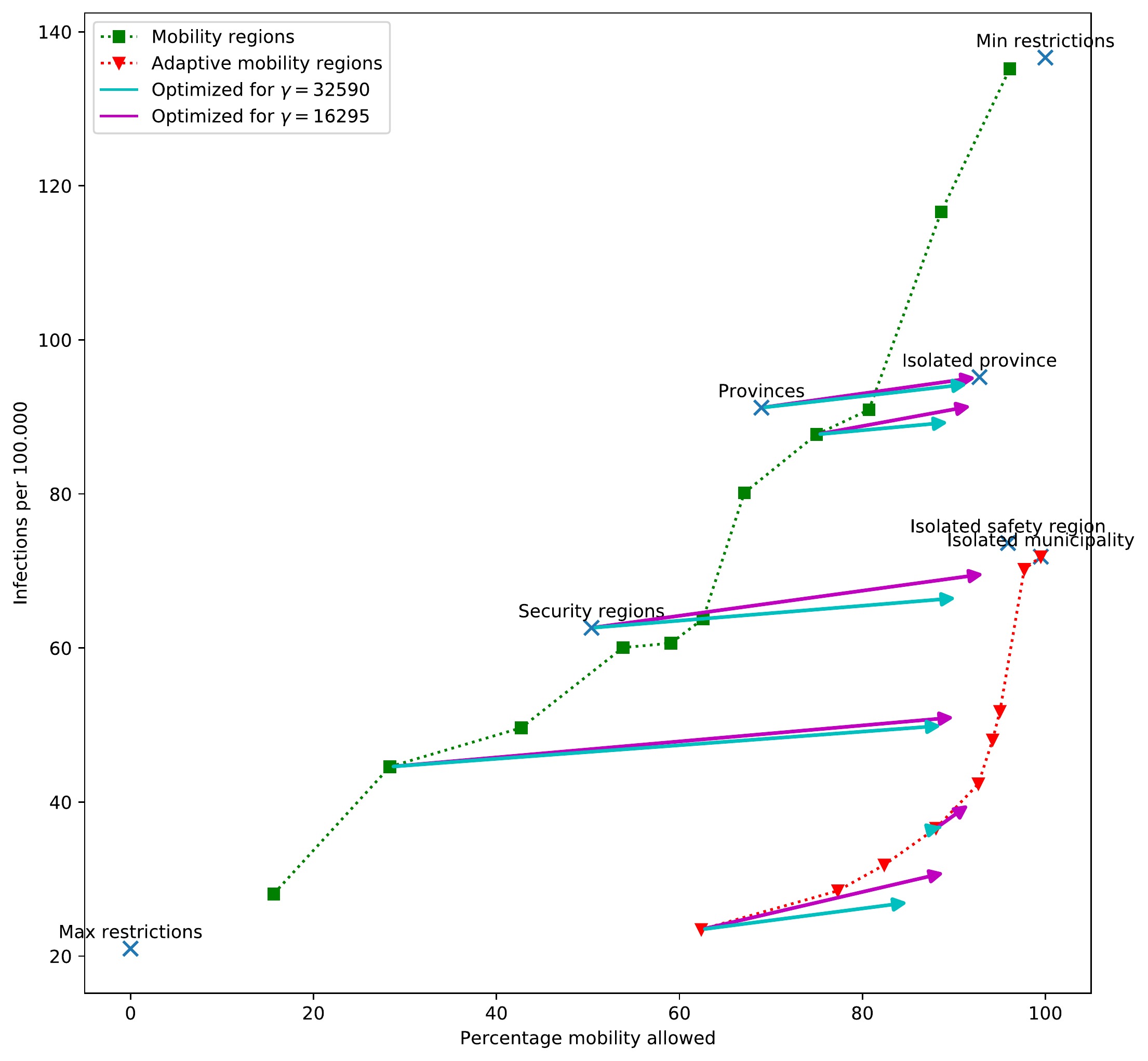}
    }
    \caption{Evaluation of different divisions for the evenly-distributed (left) versus concentrated (right) initialization. The balance of infections and mobility within an horizon of 30 days after imposing a division are shown. Blue crosses represent benchmark divisions. Green squares represent mobility regions, red triangles represent adaptive mobility regions; the parameter values $\eta$ and $\zeta$ were chosen to obtain divisions that allow various levels of mobility.   Arrows point from the starting point of the heuristic optimization to its result. We only show a single arrow for the evenly distributed initialization to avoid cluttering the figure.}
\end{figure}

\paragraph{Results for synthetic initializations.}
We consider an evenly distributed initialization where 1000 people are initially exposed. The resulting mobility and infections from each of the divisions is shown in Figure~\ref{fig:evenlydistributed}. We observe a monotonically increasing relation between infections and mobility. This may be explained by the fact that infections grow with the reproduction number, which is linear in the contact rate, which is in our model in turn linear in mobility. This trend is based on data points between $5000$ and $30000$ infections, which is a rather small difference. The exponential-like trend is unlikely to hold for long time horizons. This will depend on the speed with which local saturation in contacts starts to influence transmission potential. 

For the points shown in Figure~\ref{fig:evenlydistributed}, the division that was obtained by applying our optimization method to our mobility regions ($\eta=16$) resulted in the highest objective value for the trade-off parameter $\gamma^*$. However, this division only performs marginally better than other divisions on this part of the curve. Furthermore, with respect to a trade-off parameter of $2\cdot\gamma^*$, maximum restrictions perform best among the divisions shown.

Compared to provinces and security regions, our mobility regions heuristics can be used to tune the granularity of the divisions. This allows one to find the optimal balance on this exponential-like curve with respect to the objective function for given values of the trade-off parameter and time horizon. 
We observe no divisions which have significantly lower infection numbers in combination with as much mobility as in benchmark divisions. From this, we conclude that the amount of mobility that the division allows plays a larger role than the specific way in which this mobility is chosen.

When infections are not evenly distributed, the performances of the different divisions shift drastically. Figure~\ref{fig:concentrated} shows the performances of divisions for the concentrated initialization. In this initialization, 1000 people are initially exposed within the Dutch municipality of Uden. The mobility regions optimization does not depend on the initialization of the model so these divisions remain unchanged. The adaptive mobility regions do depend on this initialization. We see that the adaptive mobility regions significantly outperform the mobility regions and the benchmark divisions: they result in fewer infections while allowing for more mobility.
When we apply our optimization method starting from a sufficiently fine-grained division, we obtain a division that performs even better: for both choices of the trade-off parameter $\gamma^*$ and $2\cdot\gamma^*$, the division obtained from applying the optimization to the adaptive mobility regions ($\zeta=2\cdot10^7$) outperforms all the other divisions shown with respect to the objective.
This suggests that the proposed approach can provide suitable strategies for containing superspreading events, where the initial infections are highly concentrated. This holds even when we change model parameters, see Section \ref{sec-sensitivity} in the supplementary material where we perform a sensitivity analysis.

\paragraph{Historical initializations.}
Next, we evaluate divisions for an initialization based on historical data. We consider data from March 10 2020 and April 21 2020. Figure~\ref{fig:concentration} shows that their concentration values for how infections are distributed lie in between the synthetic cases. Therefore, we expect their results to also be in between the synthetic results.

On March 10 2020, the Dutch government advised all citizens of the province of Noord-Brabant to stay home. At this point, the number of reported infections was low and their distribution was far from even. April 21st was during the Dutch lock-down period, and further along in the first wave of the outbreak, where the infections were more evenly distributed. 

\begin{figure}[!ht]
 \centering
 \subfigure[March 10 2020]{
    \label{fig:historical0310}
    \centering
    \includegraphics[width=0.45\textwidth]{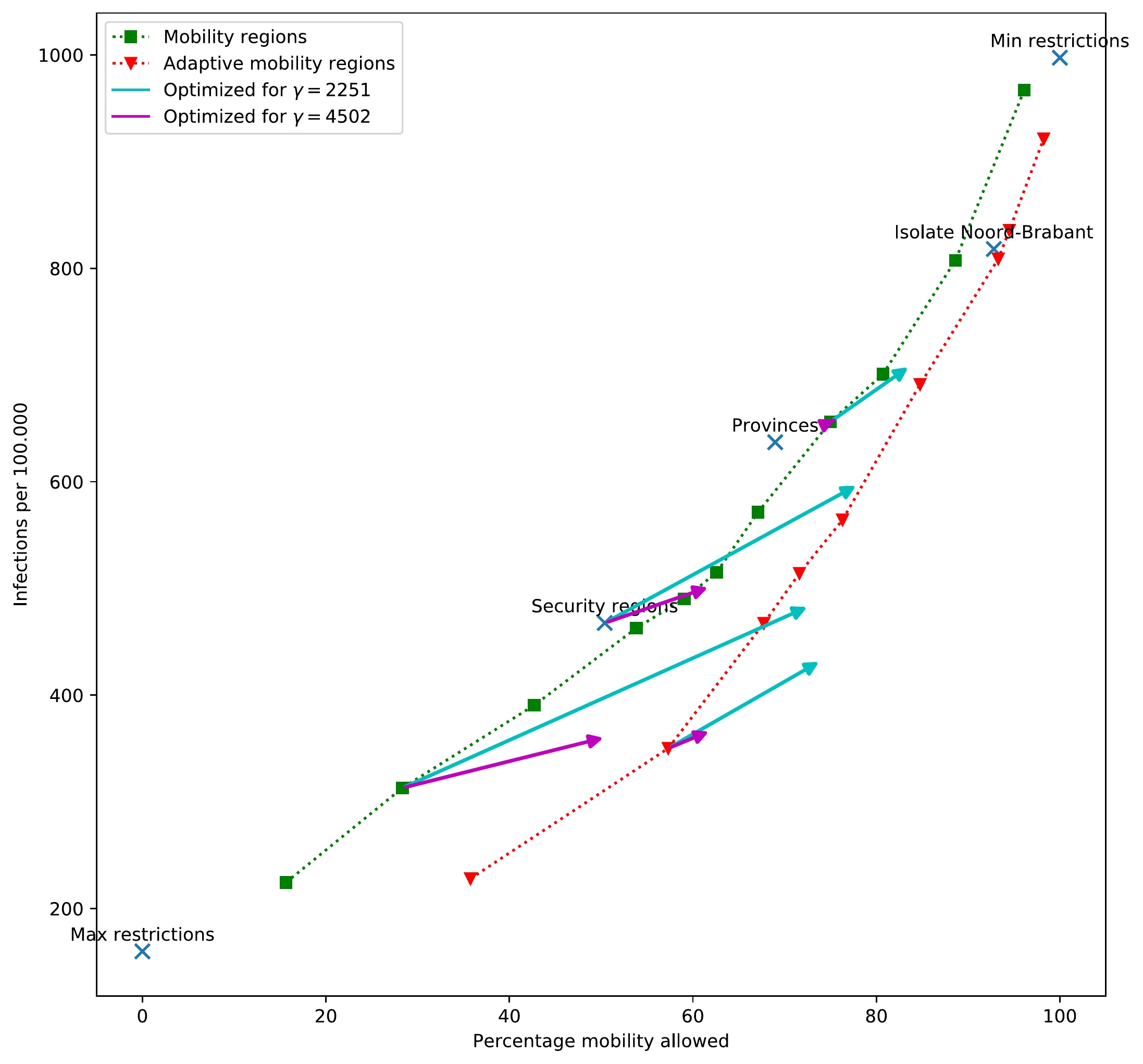}
 }
 \qquad
 \subfigure[April 21 2020]{
    \label{fig:historical0421}
    \centering
    \includegraphics[width=0.45\textwidth]{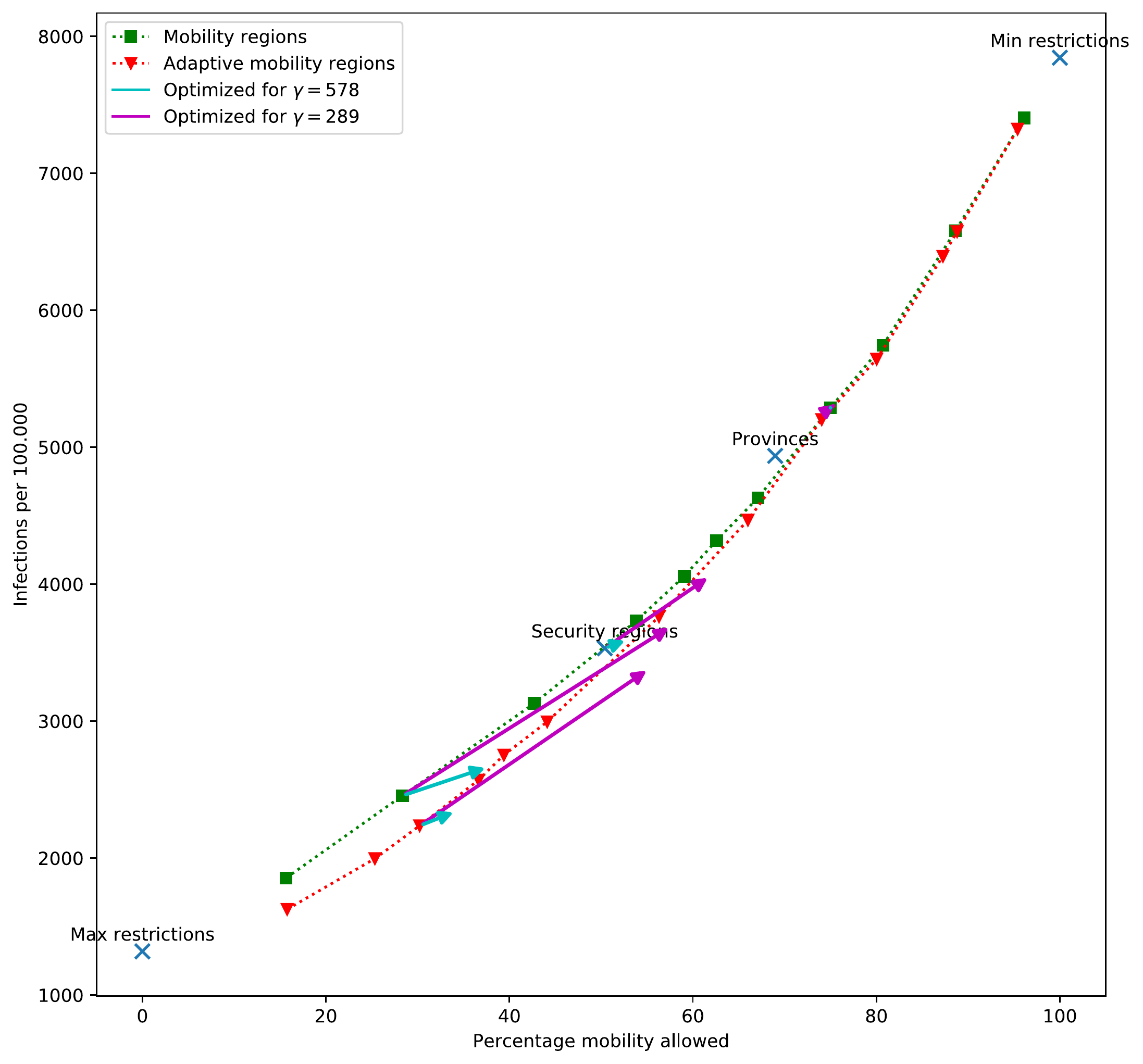}
 }
 \caption{Evaluation of divisions for initialization based on historical data of March 10 (left) versus April 21 (right) 2020. A horizon of $H=30$ days was chosen. For our mobility regions and adaptive mobility regions, the parameter values $\eta$ and $\zeta$ were chosen to obtain divisions that allow various levels of mobility.}
\end{figure}

Figure~\ref{fig:historical0310} and~\ref{fig:historical0421} show the evaluation results of different divisions. In Figure~\ref{fig:historical0310} we add a local lock-down of the province of Noord-Brabant as a benchmark division. It performs almost as good as one of the divisions from the adaptive mobility regions approach. However, at that time, the goal of the Dutch government was to suppress the virus to prevent new infections. From this perspective, only isolating Noord-Brabant is insufficient as it does not lead to significantly fewer infections than doing nothing (Min restrictions). 

Based on these two figures, we see that the findings from the synthetic initializations generalize to more realistic scenarios: when the infections have a high concentration our approach finds divisions that lead to relatively few infections while allowing for a relatively large amount of mobility. Consider the marker corresponding to the security regions division in Figure~\ref{fig:historical0310}: it can be seen that our mobility method has found divisions with 1) significantly more mobility, but a comparable amount of infections; 2) a comparable amount of mobility, but significantly fewer infections; and 3) more mobility and fewer infections. A similar conclusion can be made for the case of Figure~\ref{fig:historical0421}, though the improvements are smaller because the initialization has a lower level of concentration.

\section{Conclusion}
In a pandemic, restrictions on mobility of individuals are one of the mitigation measures available to local and national governments. Restrictions on mobility between regions will have an effect in reducing non-local transmission opportunities. The downside, however, is that restricted mobility also has potentially strong social and economic repercussions. Given this, decision makers have to reach a balance between wanted and unwanted effects when restricting mobility. Rather than impose restrictions on a national level, which could maximise unwanted effects, options need to be explored for regional measures. We have presented a method to determine an balance between infection reduction and allowed mobility. We evaluate mobility strategies that use any division of a country into regions, allowing movement within them, while disallowing movement between them. In the case of the Netherlands, we have shown that existing (administrative) divisions such as provinces and `security regions' do not reflect the mobility patterns within the country well, and therefore are not a good basis for mobility restrictions. We expect this conclusion to apply also to other countries.

We have quantified the trade-off between economy (equated here with mobility) and public health (equated here by infections) by introducing an objective function that penalizes the amount of allowed mobility by the resulting number of infections as given by Equation~\eqref{eq:tradeoff}. This trade-off introduces a parameter that can be interpreted as the number of movements that we are willing to restrict in order to prevent the occurrence of a single further infection. 

The objective function is non-linear, computationally heavy and hence infeasible to optimize exactly. Therefore, we resorted to heuristic optimization methods that are shown to produce divisions that perform well with respect to the presented objective.

As a proof of concept for the proposed methods, we have used synthetic and historical scenarios with varying concentration of infections. For each of these settings, we have compared how well the heuristics perform, and have compared the divisions obtained in this way to the benchmarks of existing divisions.
Figure~\ref{fig:evenlydistributed} shows that when the infections are evenly distributed throughout the country, the performances of all of the divisions lie close to an exponential curve. Therefore, the granularity of the division is the only relevant aspect in this case.
On the other hand, Figure~\ref{fig:concentrated} shows that when the infections are highly concentrated around one municipality, applying the optimization to adaptive mobility regions result in a divisions that significantly outperforms the others and is able to prevent more infections while allowing for more mobility.

In practice, the spread of the infections will lie between these extremes. We have introduced a formula to quantify the concentration of the infections and Figures~\ref{fig:historical0310} and~\ref{fig:historical0421} show that indeed low concentration values lead to results comparable to the case of evenly distributed infections, while higher values indeed behave similarly as the concentrated case. 
Traditional epidemiological models do not incorporate geography and therefore cannot adequately deal with situations where the infections are not evenly distributed across the country. 
In this work we have demonstrated how incorporating geometry leads to mobility-restriction strategies that are better tailored to the situation at hand. Our main conclusion is that such strategies are highly effective when the geometric spread of infections is low (so that the geometry is the main limiting factor), but it is less effective when the distribution is rather even (so that geometry is fairly irrelevant). Our main innovation is that we have proposed a method to quantify these statements.


We next discuss some possible extensions. First, we note that our model does not take any other measures into account. On the one hand, this makes the model unrealistic, while on the other hand it keeps the model simple and isolates the effects of mobility restrictions, which are our main focus. 

Secondly, the way we model mobility is especially adequate for small countries where people tend to return home after a visit to another city, such as the Netherlands. It would be interesting to use our model to analyze and compare countries that share this property. For some other countries, travelling people might stay at their destination for a period of time, instead of returning home at the end of the day. The model could be extended to allow for such behavior.

Thirdly, in our model, each region has two compartments for Tested and Untested, and each infection goes through one of these compartments. In reality, there will always be a period when the person is infected but not yet tested. Therefore, the model may be made more realistic by letting individuals transition through these compartments sequentially (possibly skipping the Tested compartment).
However, estimating the transition rates for this model can be challenging, because of its dependence on the testing policy. 

Finally, in this work we have decided to mostly base parameter choices that are relevant for the infection spread on previous studies. Alternatively, these parameters could be {\em estimated} by fitting the model to match historic data. By doing so, outcomes of the model may have a better predictive value. Currently, the numbers of infections are only used to compare with benchmark divisions. In particular, we cannot give estimates on how close these would be to the true value. Especially with historic scenarios, the fraction of reported cases and unreported cases is heavily dependent on the testing policy and availability of tests. Comparing RIVM estimates of active infections and reported infections hints that this fraction is indeed not at all constant \cite{rivm06102020}. In our model we assume this fraction to be constant in all scenarios, and it would be interesting to investigate the effect of heterogeneity, just like we now focus on the effect of heterogeneity of infections. 

 \paragraph{\bf Acknowledgments.} We thank Paul van der Schoot for stimulating discussions. The work of MG and RvdH is supported by the Netherlands Organisation for Scientific Research (NWO) through the Gravitation {\sc Networks} grant 024.002.003. The work of HH, RvdH and NL is further supported by NWO through ZonMw grants 10430022010001 and 10430 03201 0011.

\printbibliography

\appendix
\section{Supplementary material}
Here we discuss various  aspects of our approach in more detail. This appendix is organised as follows. We start in Section \ref{sec-Mezuro} by discussing the Mezuro mobility data. After this, in Section \ref{sec-comp-deter-stoch}, we discuss how our mean-field deterministic version of the SEI$_2$R$_2$-model  relates to two stochastic versions. In Section \ref{sec-init-details}, we explain the details behind our various initializations. In the final Section \ref{sec-sensitivity} we present a sensitivity analysis.

\subsection{The Mezuro mobility data}
\label{sec-Mezuro}
The Mezuro data set gives anonimized mobility data abstracted from telecommunication data. The mobility data consists of numbers of people with various origins and destinations, per day. The origins and destinations are all Dutch municipalities. Thus, a typical element could be that 211 persons travelled from the municipality called Uden with destination the municipality called Meijerijstad on March 10 2020. Here, a person is reported to have a certain destination when that person has spent at least 30 consecutive minutes in the destination municipality. Only counts that are at least 16 are being reported.

Mezuro uses the Dutch municipalities as defined 2018. At that time, there were 380 municipalities. Currently, there are 355 municipalities in the Netherlands. The average number of people living in a municipality is just over $49.000$, with the largest municipality having more than $870.000$ inhabitants (Amsterdam) and the smallest less than $1.000$ (Schiermonikoog, one of the islands in the north). In the case of the Netherlands, municipalities are on the one hand sufficiently small to make the problem of partitioning into regions interesting, on the other hand sufficiently large so that many origin-destination pairs have at least 16 people travelling between them on any given day.

The data we use as input for our model is aggregated further by taking 14-day totals and then using the daily average. We use the 14-day period running from March 1 2019 up to and including March 14 2019. By using a daily average, we lose the difference between weekdays and weekend. The period on which the average is based is deliberately chosen as before the COVID-19 outbreak. We interpret this data as being a mobility benchmark: this is what people would travel if COVID-19 were not present. Thus, if 2019 is representative for normal travel in the first half of the month March, any mobility that is less in 2020 arises through the governmental and societal measures that were imposed in response to the pandemic.


\subsection{Comparison of deterministic model with its stochastic versions}
\label{sec-comp-deter-stoch}
In this paper, we work with a {\em deterministic} SEI$_2$R$_2$ model. In this model, the spread of the infection is modeled using a discrete time evolution model. This makes that numbers are no longer integers. As a result, one obtains the well-known anomaly that when there would be $10^{-6}$, say, `people' still infected, the infection would not die out and could grow exponentially again when $\mathcal{R}_{\rm eff} > 1$.

Similarly to \cite{Li2020}, we could instead feed these non-integer values into Poisson random variables to obtain counts. We call this step {\em stochastic rounding}. The advantage is that stochastic rounding rounds most small values down to zero (a Poisson random variable with small parameter is 0 with high probability), but when there are {\em many} such random variables, occasionally one will be positive. Thus, this is a step towards a more realistic description when numbers are small.

\begin{figure}[!ht]
  \centering
  \includegraphics[width =\textwidth]{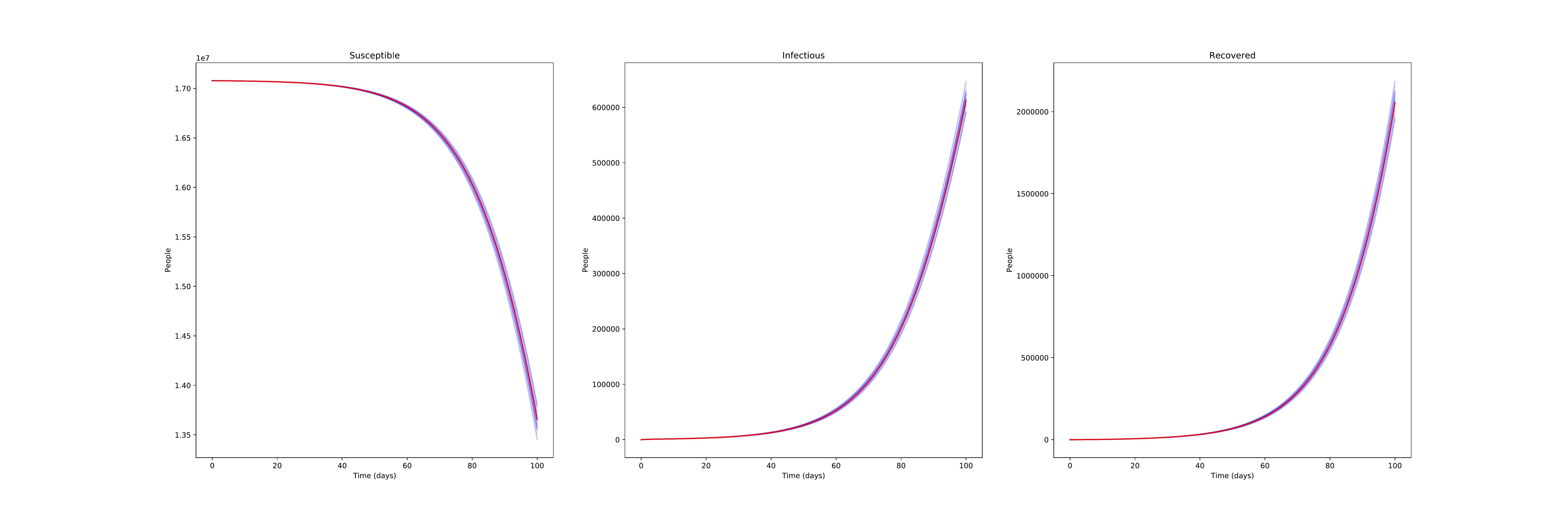}
  \caption{2000 persons are initially exposed in the municipality of Uden. The total number of people in the Netherlands who are susceptible, infectious (tested plus untested), and recovered (tested plus untested) are presented for a period of 100 days. Deterministic outcome is in red, stochastic rounding simulations are in blue.}
  \label{fig:stocashtic_2000init_100days}
\end{figure}

\begin{figure}[!ht]
  \centering
  \includegraphics[width =\textwidth]{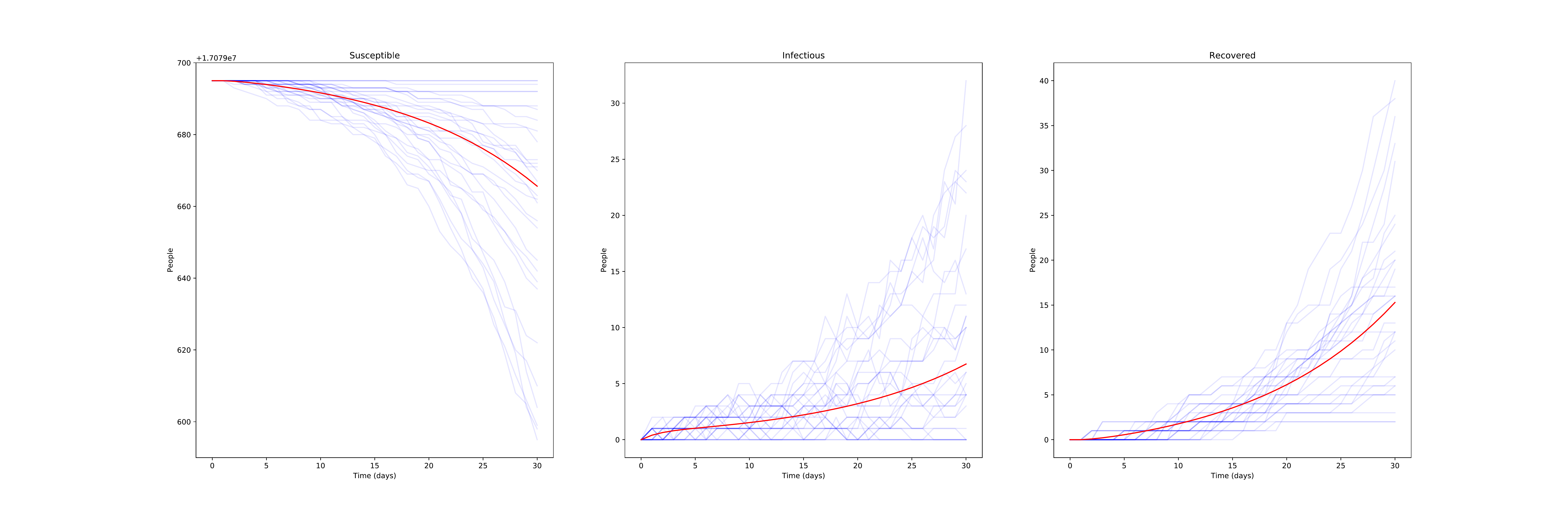}
  \caption{One person is initially exposed in the municipality of Uden. The total number of people in the Netherlands who are susceptible, infectious (tested plus untested), and recovered (tested plus untested) are presented for a period of 30 days. Deterministic outcome is in red, stochastic rounding simulations are in blue.}
  \label{fig:stocashtic_1init_30days}
\end{figure}

\begin{figure}[!ht]
  \centering
  \includegraphics[width =\textwidth]{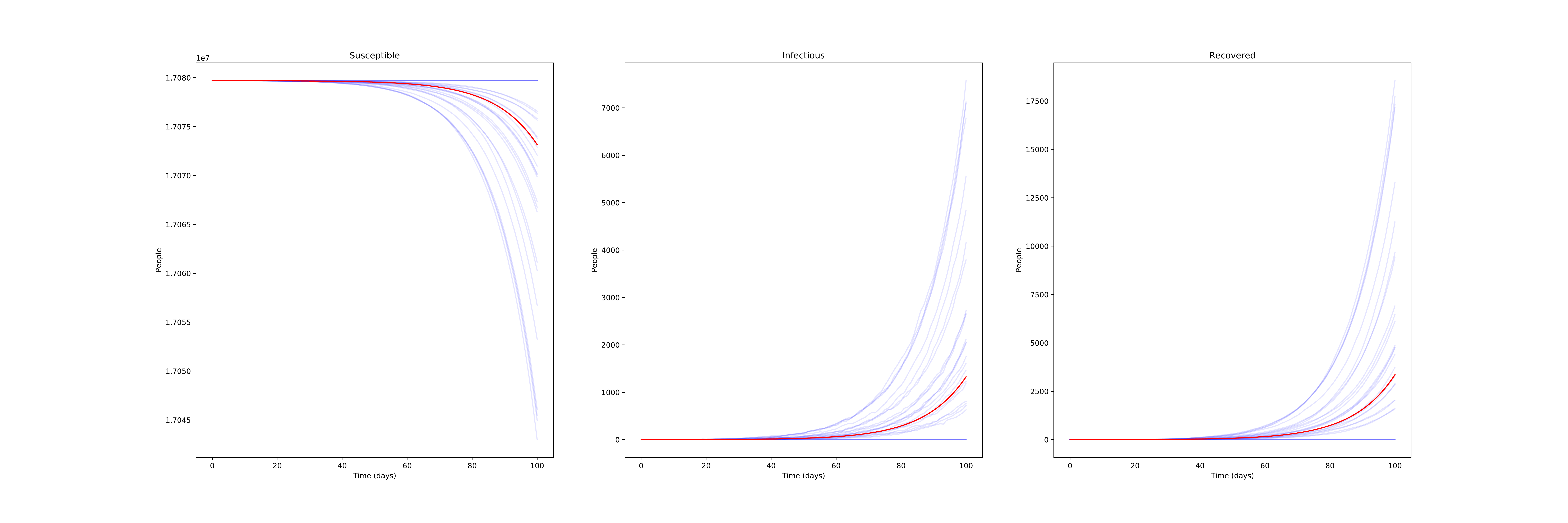}
  \caption{One person is initially exposed in the municipality of Uden. The total number of people in the Netherlands who are susceptible, infectious (tested plus untested), and recovered (tested plus untested) are presented for a period of 100 days. Deterministic outcome is in red, stochastic rounding simulations are in blue.}
  \label{fig:stocashtic_1init_100days}
\end{figure}

Simulations show that stochastic rounding has a minor effect when the numbers of infections are large, see Figure \ref{fig:stocashtic_2000init_100days}. However, in the case where an initialization is highly concentrated, the spread to neighboring regions will be strongly affected by stochastic rounding, as we might expect. In an extreme case where only 1 person is initially exposed, stochastic rounding can deviate far from the deterministic outcome, and also different simulations can deviate substantially from each other. Figure \ref{fig:stocashtic_1init_30days} and Figure \ref{fig:stocashtic_1init_100days} show results for different simulations with a time horizon of 30 days and 100 days, respectively, with one initial infected individual. There is a positive probability of the infection dying out, while, if the infection does not die out, it grows due to $\mathcal{R}_{\rm eff}>1$. 

\begin{figure}[!ht]
  \centering
  \includegraphics[width =\textwidth]{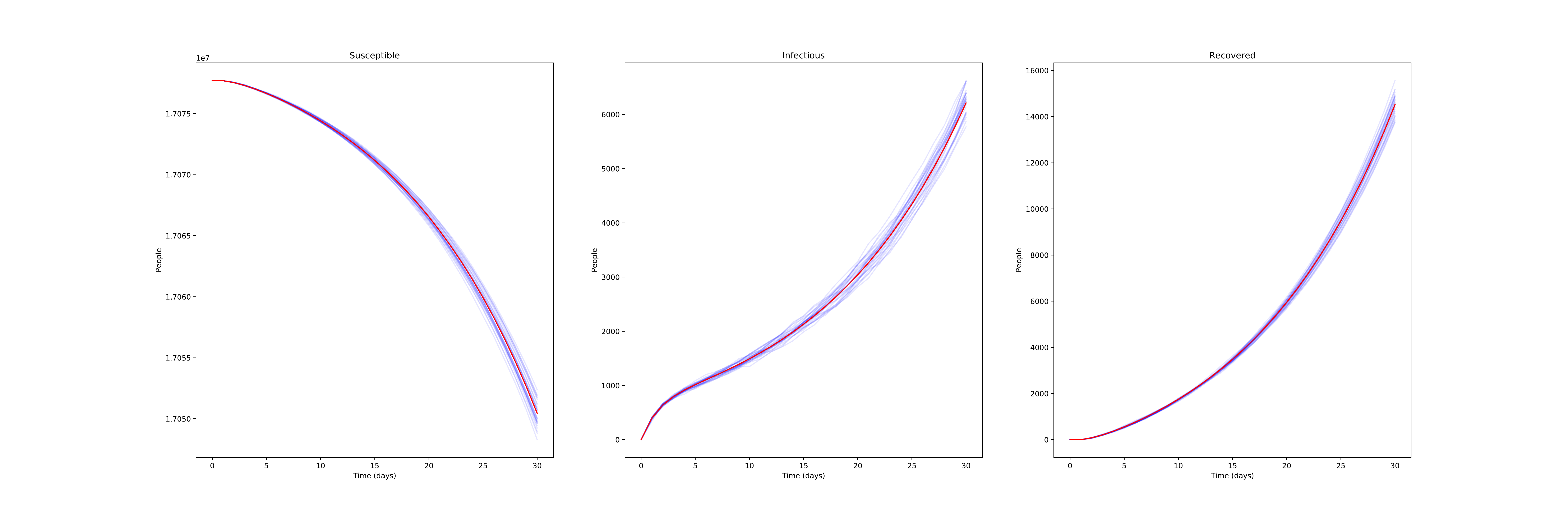}
  \caption{2000 persons are initially exposed in the municipality of Uden. The total number of people in the Netherlands who are susceptible, infectious (tested plus untested), and recovered (tested plus untested) are presented for a period of 30 days. Deterministic outcome is in red, stochastic rounding simulations are in blue.}
  \label{fig:stocashtic_2000init_300days}
\end{figure}

The spread of the virus in the version of the model with stochastic rounding can be much reduced compared to the version without rounding. This means that we have to interpret our results with caution when there are many counts that are quite small.  

\begin{figure}[!ht]
  \centering
  \includegraphics[width =\textwidth]{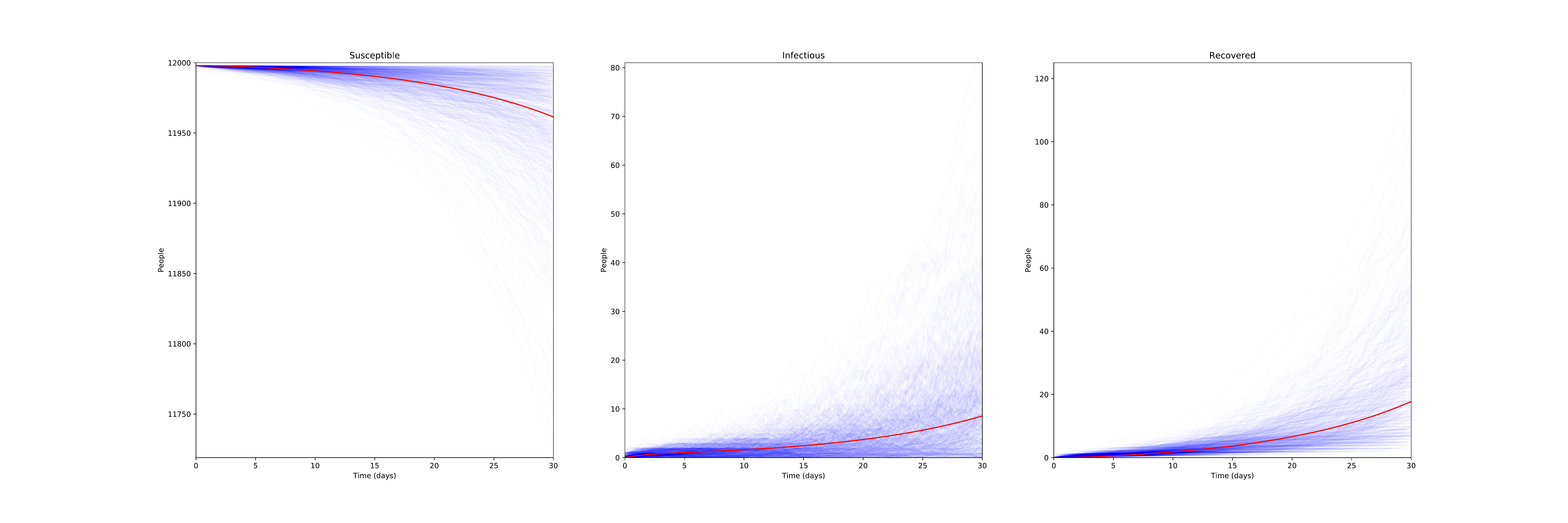}
  \caption{One person is initially exposed. Total number of people who are susceptible, infectious (tested plus untested), and recovered (tested plus untested) are presented for a period of 30 days. Deterministic outcome is in red, stochastic simulations are in blue.}
  \label{fig:fullstocashtic_1init_30days}
\end{figure}

\begin{figure}[!ht]
  \centering
  \includegraphics[width =\textwidth]{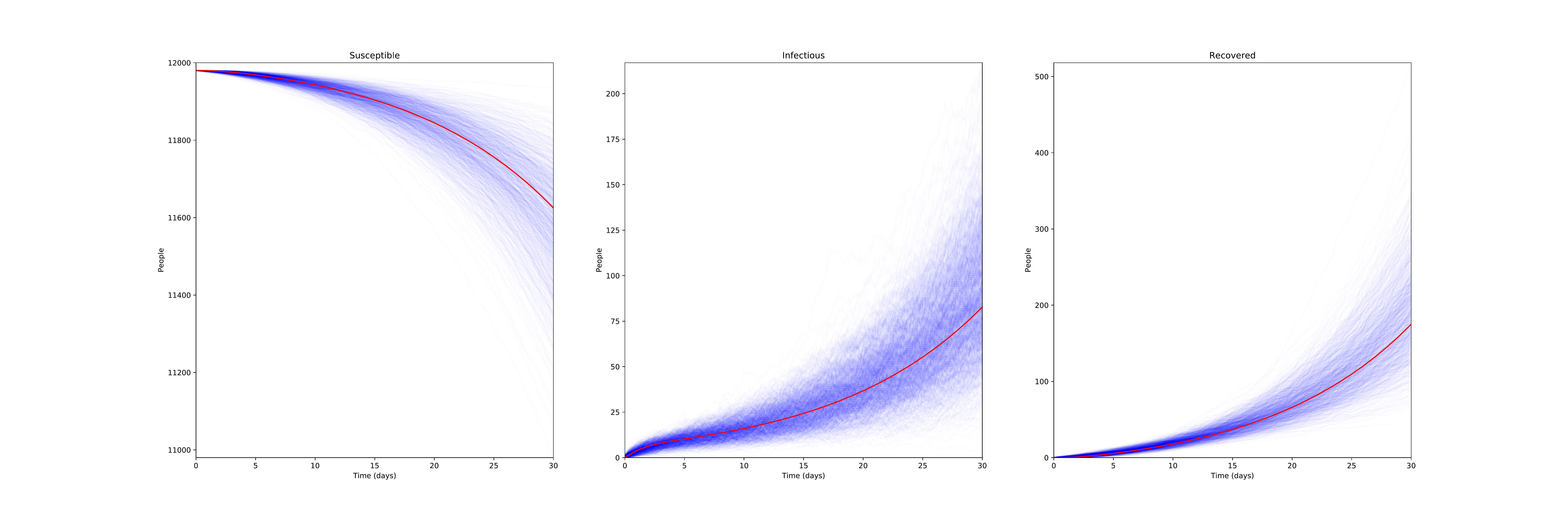}
  \caption{Ten people are initially exposed. Total number of people who are susceptible, infectious (tested plus untested), and recovered (tested plus untested) are presented for a period of 30 days. Deterministic outcome is in red, stochastic simulations are in blue.}
  \label{fig:fullstocashtic_10init_30days}
\end{figure}

The spread of an epidemic is inherently a random process, where the stochasticity plays a much more profound role than what is represented by the stochastic rounding. We have implemented a particle-based (event-based) model, where we simulate our SEI$_2$R$_2$ for two regions, to see whether there are big differences between the particle-based version of the model and our deterministic version. Simulations show that when the number of infections is high, the deterministic version performs well, while when this number is low, the predictions are much less accurate. Indeed, when the number of infections is small, the stochastic SEI$_2$R$_2$-model has a positive probability of dying out, while the deterministic model will not. 
See Figure \ref{fig:fullstocashtic_1init_30days} and Figure \ref{fig:fullstocashtic_10init_30days} for an example of such a simulation. In these figures we simulate a system of two regions with 8000 and 4000 inhabitants. In Figure \ref{fig:fullstocashtic_1init_30days}, one person is initially exposed in both regions, while in Figure \ref{fig:fullstocashtic_10init_30days}, 10 people are initially exposed in both regions.

\subsection{Details on the precise initializations}
\label{sec-init-details}
In this section, we discuss the details of the different initializations as discussed in the paper.

\paragraph{Evenly-distributed initialization.} A fixed fraction $r$ of people in all municipalities is set to exposed, i.e.,
\[
S_i\left(t_0\right)=\left(1-r\right)N_i,\qquad
E_i\left(t_0\right)=rN_i.
\]
Note that the number of people who are susceptible does not need to be integer-valued in a deterministic simulation. We initialize by rounding all values to the closest integer. 
After initializing, we simulate for $10$ days to mimic that the virus has spread to other municipalities before it is found and people are tested. The resulting state at $t= t_0+10$ can then be used as an input for the adaptive mobility regions approach and also as a starting point for evaluating the performance of different divisions.

\paragraph{Concentrated initialization.}  We look at the situation after a superspreader event (subscript sse), where the number of exposed people after such an event equals $E_\text{sse}$, e.g. a superspreader event at a meat factory in the municipality of Groenlo where eventually $68$ inhabitants were tested positively . We initialize by only exposing people in one municipality, other municipalities are $100\%$ susceptible and assume that at that point in time there are no other active cases. The municipality where the superspreader event has happened is referred to as superspreader municipality (subscript ssm). We initialize as
\[
S_\text{ssm}\left(t_0\right)=N_\text{ssm}-E_\text{sse},\qquad
E_\text{ssm}\left(t_0\right)=\ E_\text{sse}.
\]
Similarly in the case of evenly-distributed initializations, we initialize by rounding all values to the closest integer. 
After initializing, we simulate for $10$ days to mimic that the virus has spread to other municipalities before it is found and people are tested. The resulting state at $t= t_0+10$ is then used as input for our approach.

\paragraph{Initialization based on historic data.} For an initialization based on historic situations, RIVM data is used (RIVM is the Dutch institute of public health). The historic data have a distribution that is neither even nor concentrated. RIVM reports the cumulative number of infections for each municipality. The number of active known infections at time $t$, i.e. tested individuals $I^T_i(t)$, is equal to the reported infections at time $t$, $I_i^\text{RIVM}\left(t\right)$, minus the reported infections one infectious period ago $I_i^\text{RIVM}\left(t-\omega\right)$. We assume that a this is only a $a$ of the actual number of active infected individuals. A fraction (1-a) is not tested by RIVM and unknown to RIVM. We initialize as follows:
\[
I_i^\text{T}\left(t\right)= \left(I_i^\text{RIVM}\left(t\right)-\ I_i^\text{RIVM}\left(t-\omega\right)\right).
\]
Untested active infections $I_i^\text{U}\left(t\right)$ are initialized similarly,
\[
I_i^\text{U}\left(t\right)= \frac{1-a}{a}\left(I_i^\text{RIVM}\left(t\right)-\ I_i^\text{RIVM}\left(t-\omega\right)\right).
\]
Recovered people are initialized as
\[
R_i^\text{T}\left(t\right)= I_i^\text{RIVM}\left(t-\omega\right),\quad
R_i^\text{U}\left(t\right)=\frac{1-a}{a}I_i^\text{RIVM}\left(t-\omega\right).
\]
Exposed people are people who will become infectious after the latency period has ended. The number of people exposed at time $t$ is equal to the infections one latency period from now $I^T_i\left(t+\nu\right)+I^U_i\left(t+\nu\right)$, minus the infections at time $t$, $I^T_i\left(t\right)+I^U_i\left(t\right)$. In terms of reported cases by RIVM, we can rewrite \[I^T_i\left(t\right)+I^U_i\left(t\right) = I_i^\text{RIVM}\left(t\right)+\frac{1-a}{a}I_i^\text{RIVM}\left(t\right)=\frac{1}{a}I_i^\text{RIVM}\left(t\right)\].
Then, exposed are initialized as 
\[
E_i\left(t\right)=\frac{1}{a}\left(I_i^\text{RIVM}\left(t+\nu\right)-\ I_i^\text{RIVM}\left(t\right)\right).
\]
Susceptible people are those that are not within one of the other compartments, i.e.,
\[
S_i\left(t\right)=N_i-E_i\left(t\right)-I_i^\text{T}\left(t\right)-I_i^\text{U}\left(t\right)-R_i^\text{T}\left(t\right)-R_i^\text{U}\left(t\right).
\]
This initial state can be used as input for the adaptive mobility regions or as a starting point for evaluating the performance of different divisions.
\medskip

\paragraph{Concentration values of different initializations.}
Using the initialization methods as described above, we are able to vary the initial concentration of infected individuals. We can simulate cases where it is evenly distributed over the Netherlands, but also where it is concentrated in only a single municipality. By using historical data, we end up in a situation which is in-between the two extremes, see also Figure \ref{fig:concentration}.

\subsection{Sensitivity Analysis}
\label{sec-sensitivity}
In this section, we briefly show the sensitivity of our results with respect to the different input parameters. Also, we discuss the reproduction number.

\begin{figure}[!h]
 \centering
 \subfigure[$a=0.01$]{
        \label{fig:concentrated_a_0.01}
        \centering
        \includegraphics[width=0.45\textwidth]{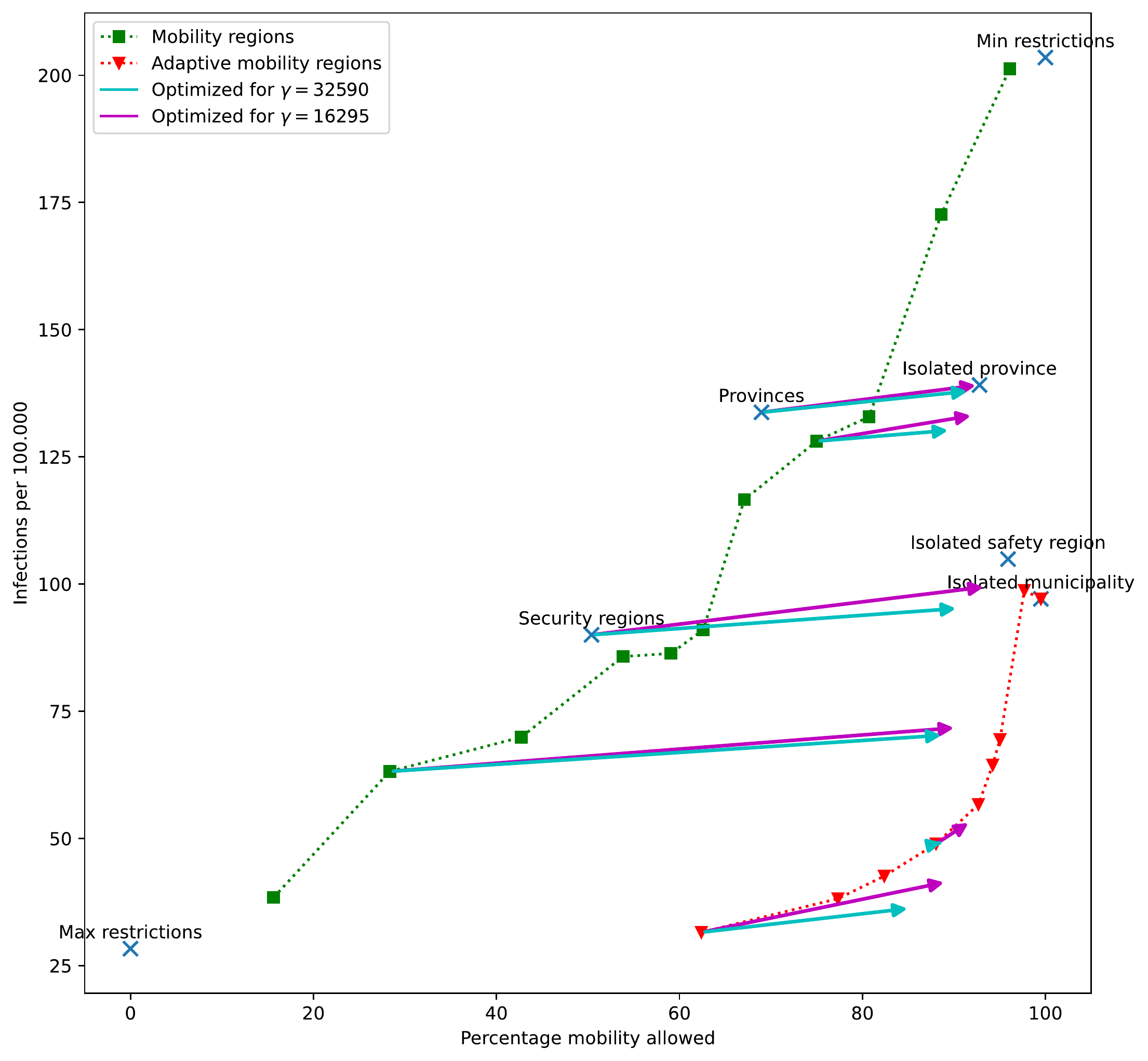}
    }%
    \qquad
    \subfigure[$a=1$]{
        \label{fig:concentrated_a_1}
        \centering
        \includegraphics[width=0.45\textwidth]{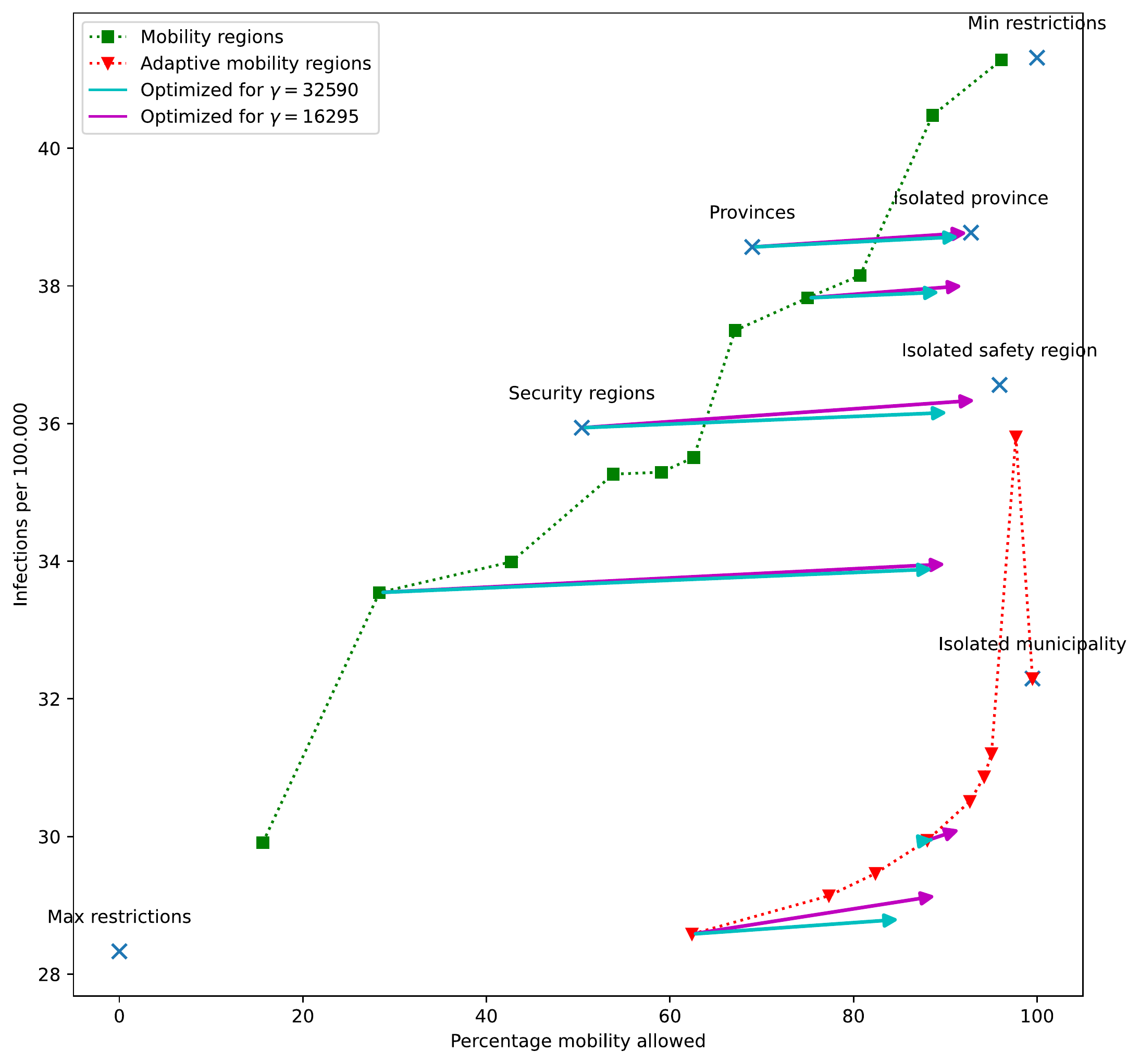}
    }
    \caption{Analyzing robustness of Figure \ref{fig:concentrated} with respect to the fraction of tested people.}
    \label{fig:robust_a}
\end{figure}

\begin{figure}[!ht]
 \centering
 \subfigure[$p=0.1$]{
        \label{fig:concentrated_p_0.1}
        \centering
        \includegraphics[width=0.45\textwidth]{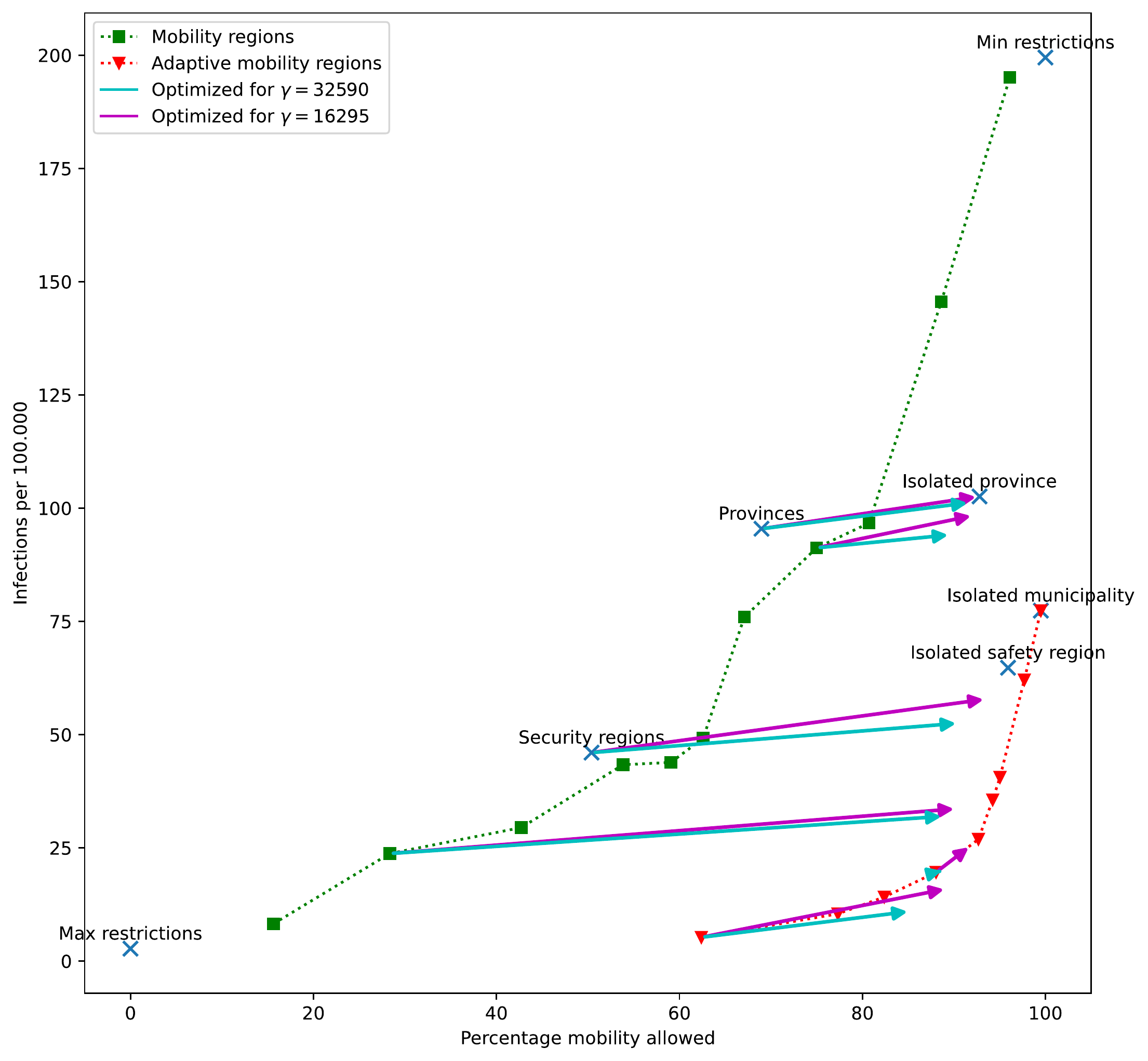}
    }%
    \qquad
    \subfigure[$p=0.9$]{
        \label{fig:concentrated_p_0.9}
        \centering
        \includegraphics[width=0.45\textwidth]{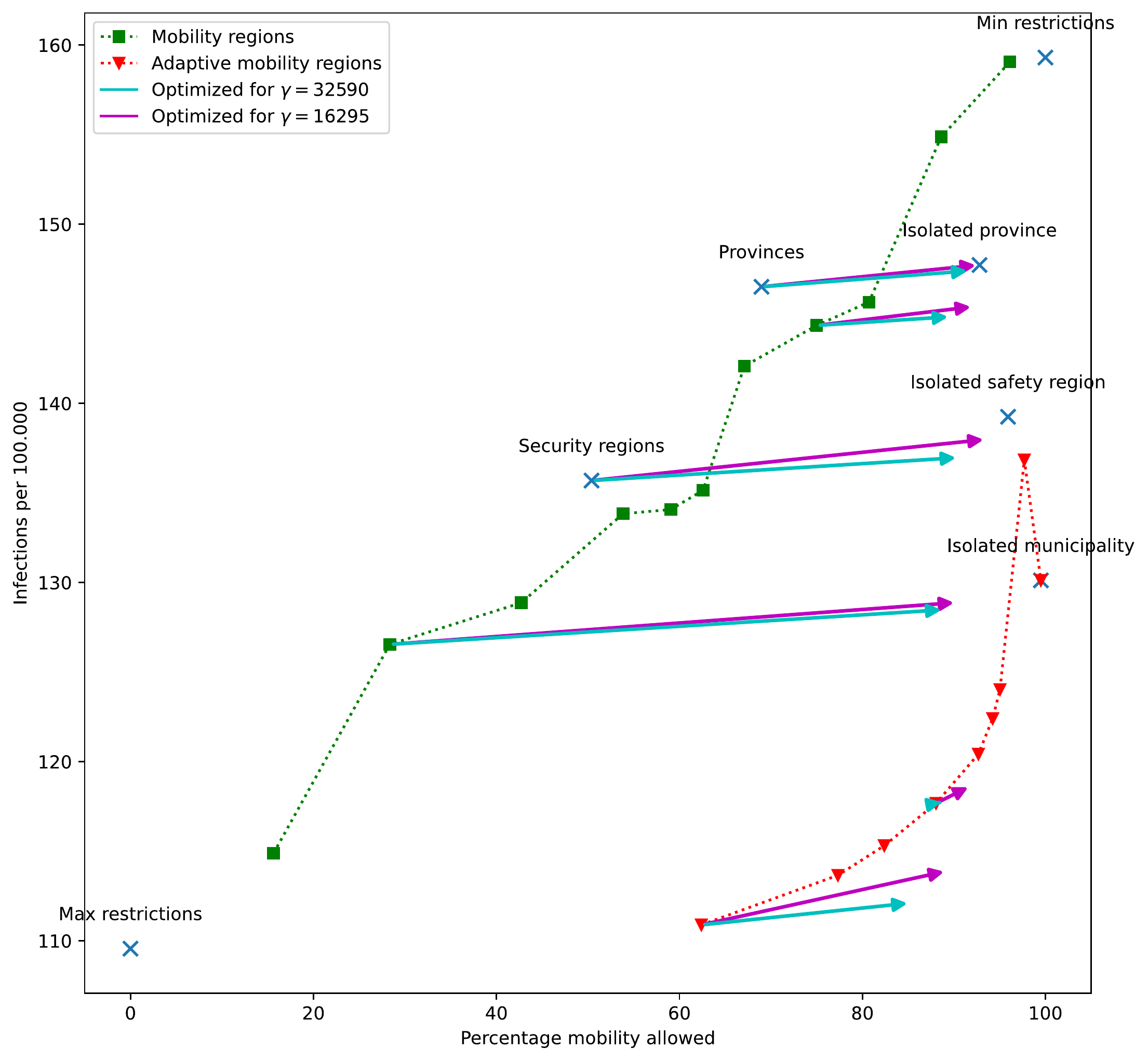}
    }
    \caption{Analyzing robustness of Figure \ref{fig:concentrated} with respect to the fraction of local contacts.}
    \label{fig:robust_p}
\end{figure}

\begin{figure}[!ht]
 \centering
 \subfigure[$\mathcal{R}_0=1$]{
        \label{fig:concentrated_R_1}
        \centering
        \includegraphics[width=0.45\textwidth]{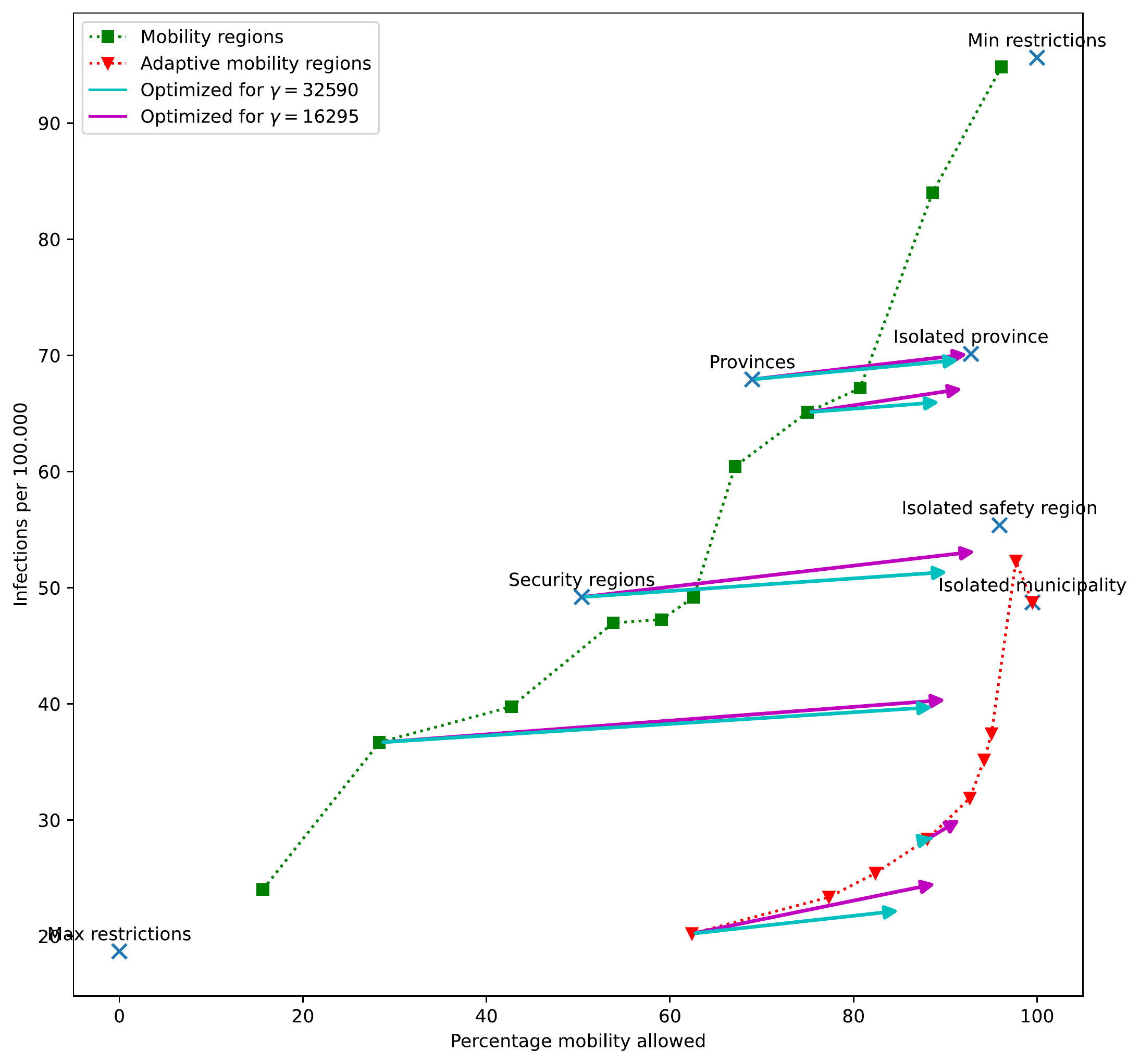}
    }%
    \qquad
    \subfigure[$\mathcal{R}_0=2.5$]{
        \label{fig:concentrated_R_2.5}
        \centering
        \includegraphics[width=0.45\textwidth]{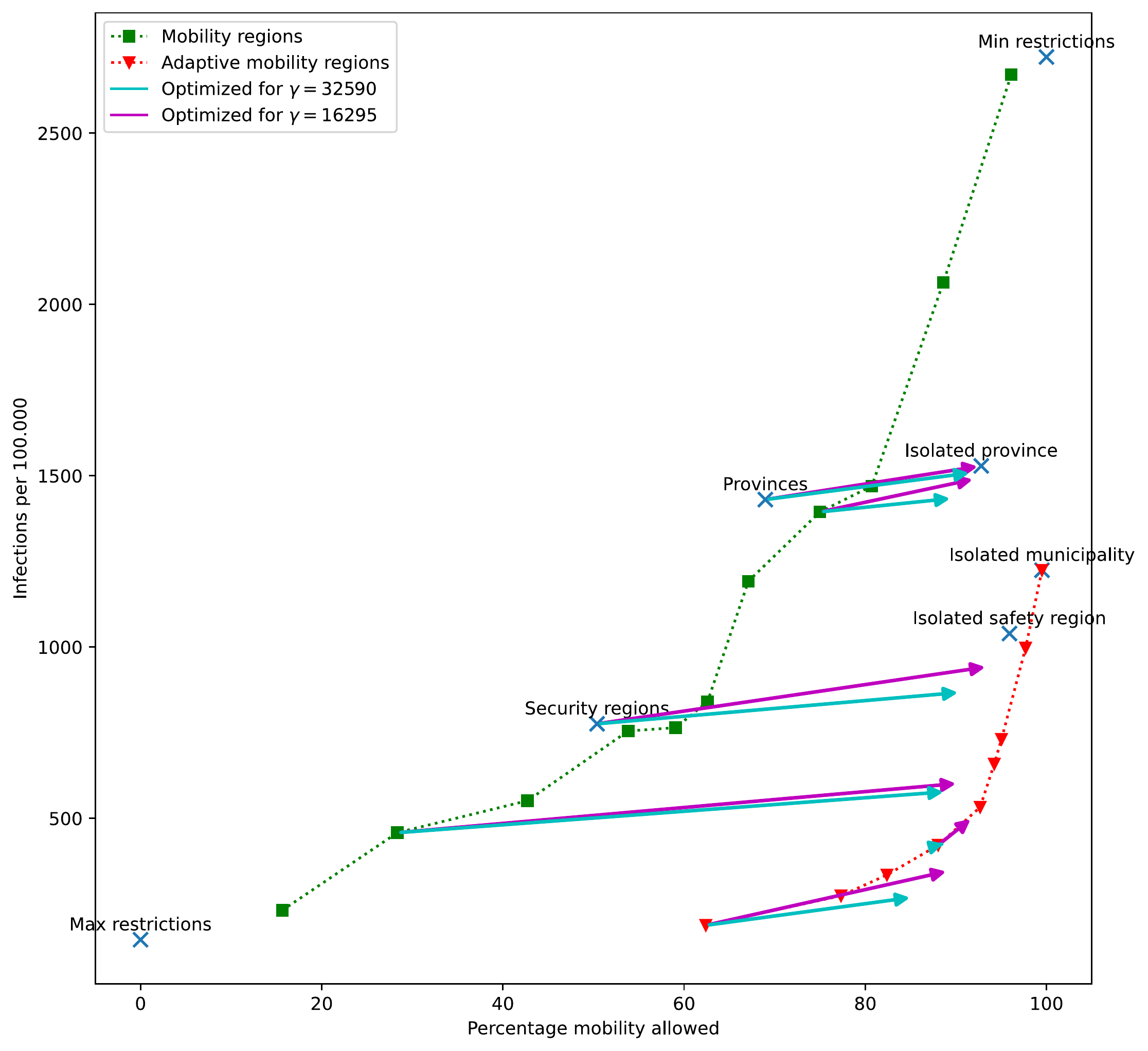}
    }
    \caption{Analyzing robustness of Figure \ref{fig:concentrated} with respect to the effective reproduction number.}
    \label{fig:robust_R}
\end{figure}

\begin{figure}[!ht]
 \centering
 \subfigure[$\omega=3$ days]{
        \label{fig:concentrated_w_1}
        \centering
        \includegraphics[width=0.45\textwidth]{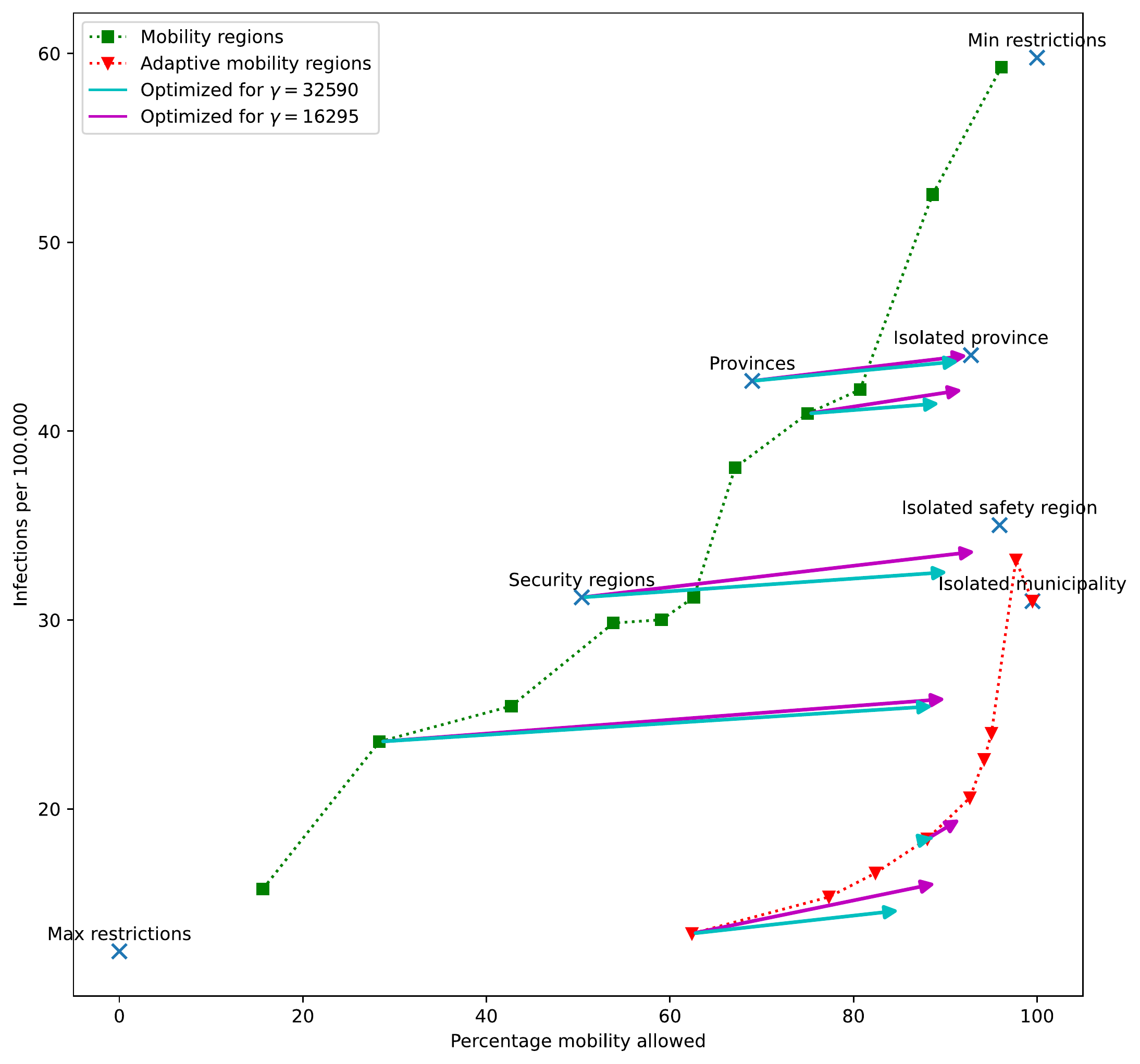}
    }%
    \qquad
    \subfigure[$\omega=7.5$ days]{
        \label{fig:concentrated_w_2.5}
        \centering
        \includegraphics[width=0.45\textwidth]{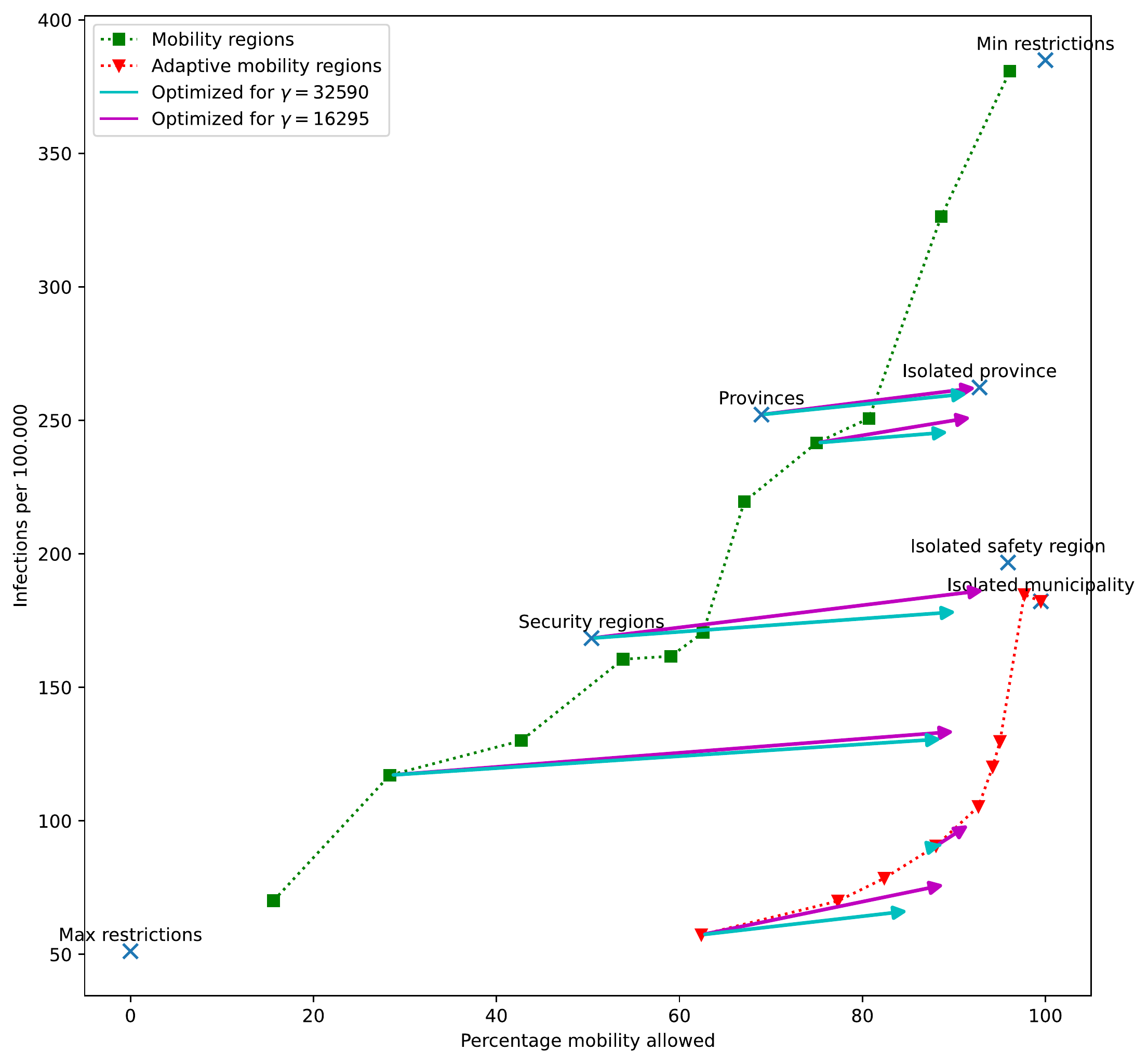}
        \label{fig:robust_w}
    }
    \caption{Analyzing robustness of Figure \ref{fig:concentrated} with respect to the infectious period.}
\end{figure}

\begin{figure}[!ht]
 \centering
 \subfigure[$\nu=3$ days]{
        \label{fig:concentrated_v_1}
        \centering
        \includegraphics[width=0.45\textwidth]{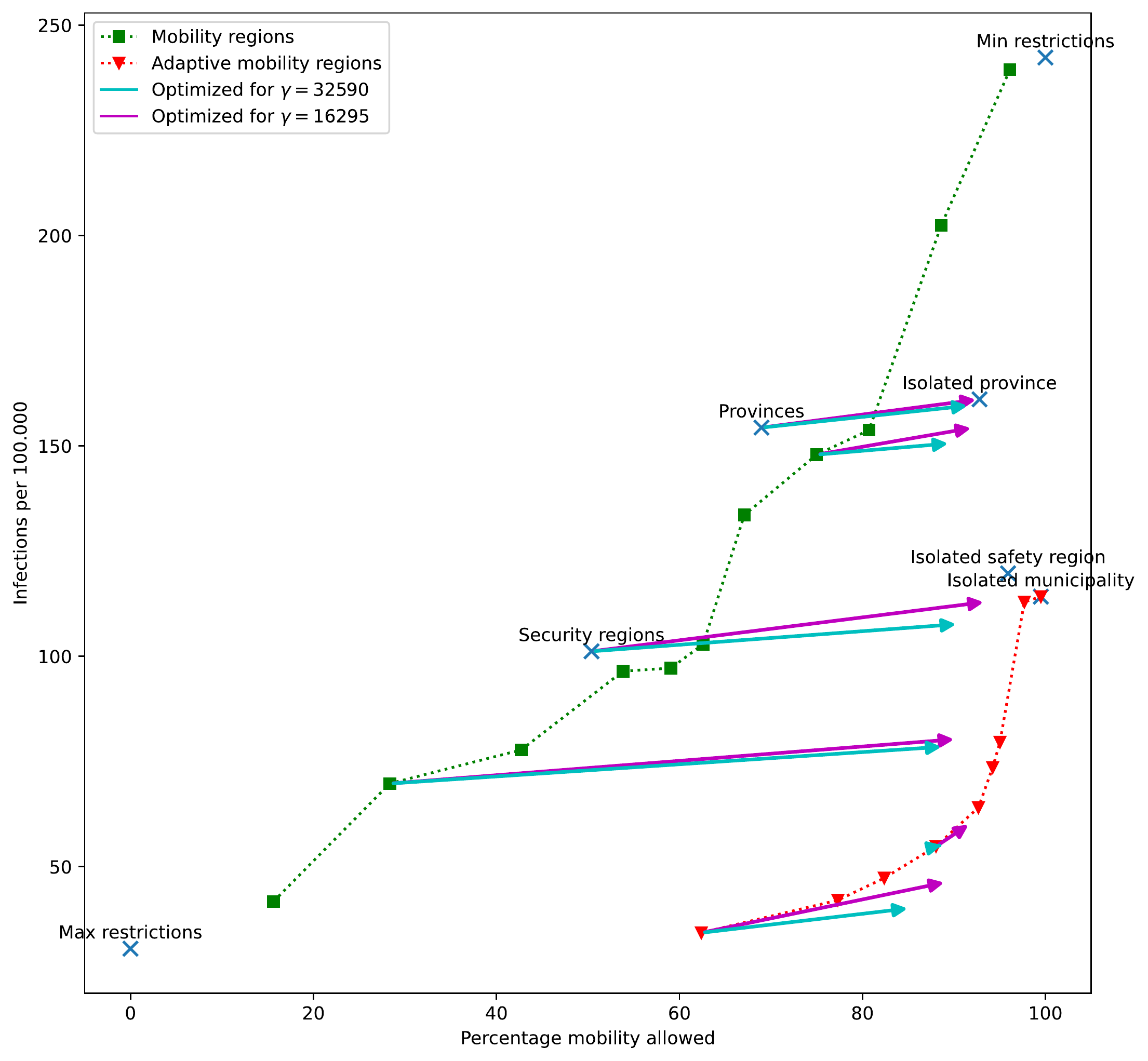}
    }%
    \qquad
    \subfigure[$\nu=7.5$ days]{
        \label{fig:concentrated_v_2.5}
        \centering
        \includegraphics[width=0.45\textwidth]{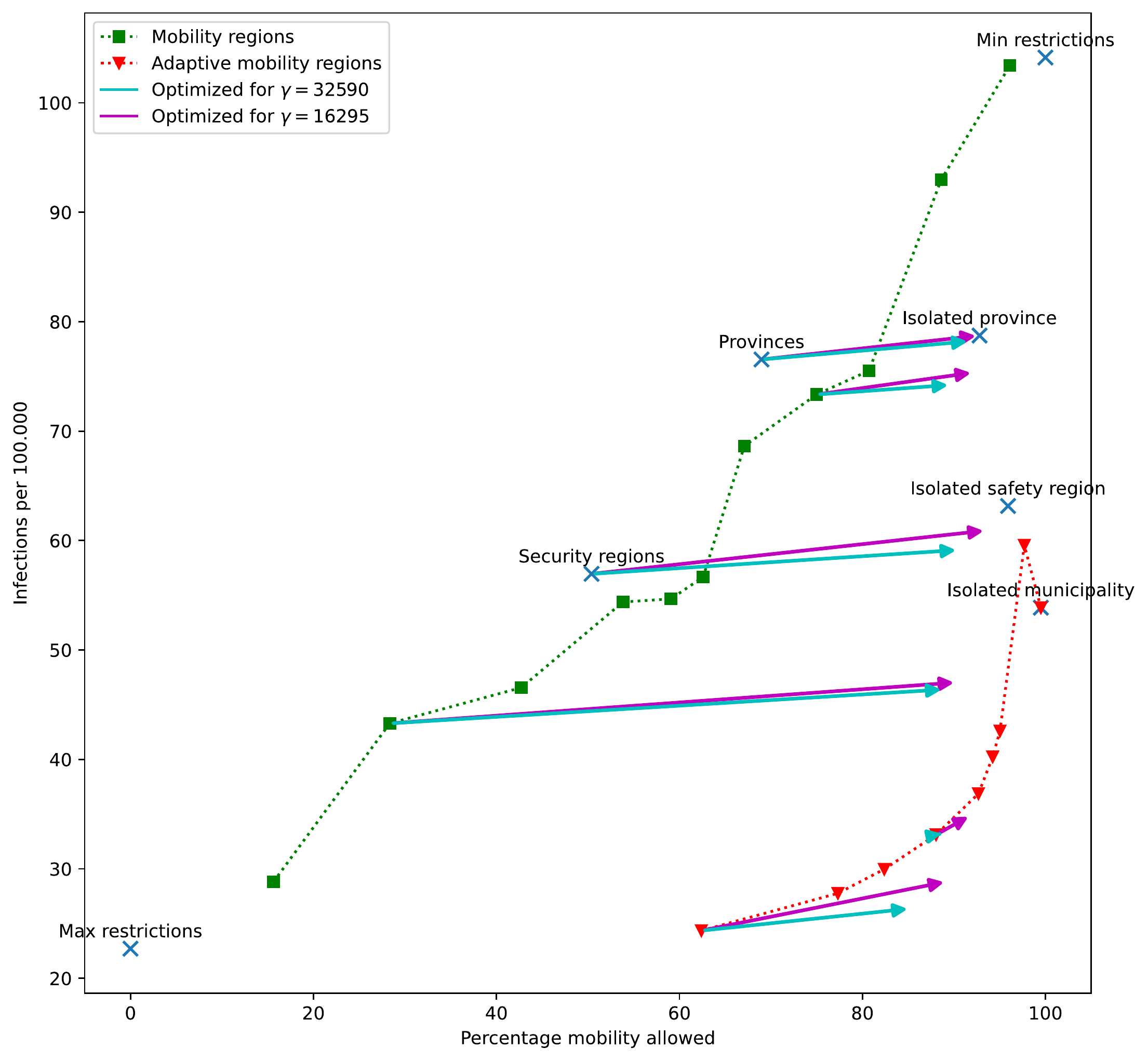}
    }
    \caption{Analyzing robustness of Figure \ref{fig:concentrated} with respect to the latent period.}
    \label{fig:robust_v}
\end{figure}

\paragraph{Robustness of results.} We analyze the sensitivity of the results to the model parameters by varying parameters one-at-a-time, using the values from Table \ref{tab:sensitivity_analysis}. The results in Figures \ref{fig:evenlydistributed}, \ref{fig:concentrated}, \ref{fig:historical0310}, \ref{fig:historical0421} are recalculated with different parameters of the model and  are shown in Figures \ref{fig:robust_a}, \ref{fig:robust_p}, \ref{fig:robust_R}, \ref{fig:robust_R}, \ref{fig:robust_w}, and \ref{fig:robust_v}.

\begin{table}[]
\centering
\label{tab:sensitivity_analysis}
\begin{tabular}{|l|llll|}
\hline
Name                      & Variable & Base value & Lower value & Upper value \\
\hline
Fraction tested           & $a$        & 1/15       & 0.01        & 1           \\
Fraction local contacts   & $p$        & 0.5        & 0.1         & 0.9         \\
Effective reproduction number & $R_{\rm eff}$    & 1.25       & 1           & 2.5         \\
Infectious period         & $\omega$ & 5          & 3           & 7.5         \\
Latent period              & $\nu$    & 5          & 3           & 7.5\\
\hline
\end{tabular}
\caption{Ranges of parameters for our sensitivity analysis}
\end{table}

Differences are most noticeable for simulations where the locations of the infectious people are concentrated. But even then, our conclusions are fairly robust against changing parameters. Infection counts on  the $y$-axis do differ a lot, but the shape of the curves stay roughly the same, and thus, our conclusions also remain unchanged. This effect is clearly visible when we change the effective reproduction number to 1 and to 2.5. Then, infections after 30 days are at 9000 or 50,000, but the relative performance of different divisions is similar. 

\paragraph{Reproduction number.}
We calculate the reproduction number as the dominant eigenvalue of the Next Generation Matrix (NGM) \cite{diekmann2010construction,diekmann1990definition}. The basic reproduction number $\mathcal{R}_0$ is a measure for the expected offspring of a single infectious person in a completely susceptible population, i.e. when $S\approx N$, in an uncontrolled situation. We focus on the effective reproduction number in a situation where testing is active as a strategy (as quantified by $a$) of control and restrictions on non-local mobility (as quantified by the value of the fraction local contacts, $p$). Stronger mobility restrictions translate into higher values of $p$.

An infectious person in region $i$ can spread the disease in three different ways. Firstly, local spread occurs within region $i$. Secondly, infected individuals travel to a different region $j$ and infect susceptible people living there. Thirdly, susceptibles from region $j$ travel to region $i$ and can get infected there. The latter two are only possible when the infectious person is allowed to travel, in other words, he/she is untested, which happens with probability $1-a$.

The NGM elements are constructed by a direct epidemiological reasoning. The NGM is represented by matrix $\textbf{K}$. The element $K_{ij}$ is the expected number of new infections in region $i$ caused by an infectious person from region $j$. Hence, $K_{ii}$ are local infections and $K_{ij},\ i\neq j$ are infections related to mobility.

The diagonal elements of the NMG are equal to 
\[K_{ii}=\beta_{\rm loc}w,\] 
and off-diagonal elements are equal to
\[K_{ij}=\beta_{\rm mob}w(1-a)\frac{M_{ji}}{N_i},\quad i\neq j.\]
For $\beta_{\rm loc}=0.165$, $\beta_{\rm mob}=0.141$, $\alpha=\frac{1}{15}$, and $w=5$, the effective reproduction number equals $R_{\rm eff}=1.25$. We can bring 2 parameters of the model in front of the NGM, namely $\omega$ and $\varepsilon$. We then find
\[\frac{K_{ii}}{\varepsilon\omega}=c_{\rm loc}=pc,\] 
and 
\[\frac{K_{ij}}{\varepsilon\omega}=c_{\rm mob}(1-a)\frac{M_{ji}}{N_i} = (1-p)c\frac{N}{2\sum_{i,j\in\mathcal{A}}M_{ij}}(1-a)\frac{M_{ji}}{N_i},\quad i\neq j.\]

\begin{figure}
    \centering
    \includegraphics[width=0.85\textwidth]{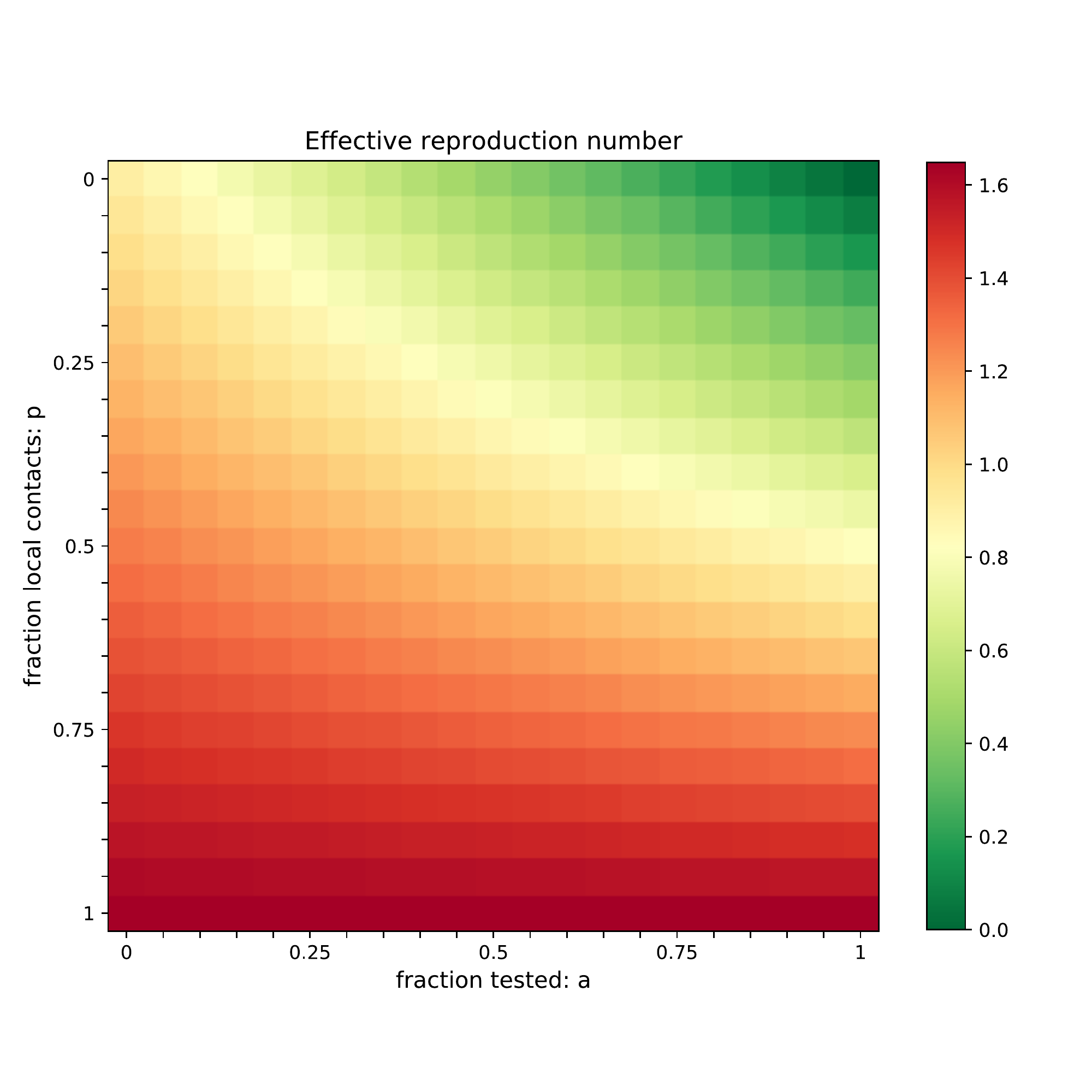}
    \caption{Effective reproduction number as a function of the fraction of people who is tested $a$ and the fraction of contacts which a person has within its own municipality $p$. We use $\omega=5$ days, $c=13.85$ contacts per day, and a transmission probability of $\varepsilon= 0.0238$.}
    \label{fig:reproduction_number}
\end{figure}

Here we can also bring $c$ in front of the NGM, so that we end up with elements depending only on $a$ and $p$, and some system constants as population sizes and mobility numbers. We calculate $R_{\rm eff}$ as a function of $a$ and $p$, see Figure \ref{fig:reproduction_number}.

This figures provides insights in what the different combinations of the values for fraction tested $a$ and local contacts $p$ can be effective as control measure, i.e. force the effective reproduction te be lower than 1. For example, if one would be able to find close to all infectious individuals by testing $\left(a\approx1 \right)$, in combination with people making 50\% of their contacts locally $\left( p=0.5\right)$, the Effective reproduction number is below 1 without mobility restrictions. We also see that if we do not test enough, the reproduction number is likely to be larger than 1 and additional mobility restrictions are necessary for containment.

\end{document}